# Small moment antiferromagnetic ordering in single crystalline La$_2$Ni$_7$.

R. A. Ribeiro,[1,2] S. L. Bud'ko,[1,2] L. Xiang,[1,2] D. H. Ryan,[1,3] P. C. Canfield[1,2]

[1] *Ames Laboratory, Iowa State University, Ames, Iowa, 50011, U.S.A.*

[2] *Department of Physics and Astronomy, Iowa State University, Ames, Iowa, 50011, U.S.A.*

[3] *The Centre for the Physics of Materials and the Physics Department, McGill University, Montreal, H3A 2T8, Canada.*

**Abstract**

Single crystals of La$_2$Ni$_7$ have been grown out of a binary, La-Ni melt. Temperature dependent, zero magnetic field, specific heat, electrical resistivity, and low field magnetization measurements indicate that there is a series of antiferromagnetic phase transitions at $T_1 = 61.0 \pm 0.2$ K, $T_2 = 56.5 \pm 0.2$ K and $T_3 = 42.2 \pm 0.2$ K. The three specific heat anomalies found at these temperatures qualitatively have very small entropy changes associated with them and the anisotropic M(H) data saturate at ~ 0.12 $\mu_B$/Ni; both observations strongly suggesting the AFM order is associated with very small, itinerant, moments. Anisotropic, H$_{\parallel c}$ and H$_{\perp c}$, $\rho$(H) and M(H) isotherms as well as constant field, $\rho$(T) and M(T) sweeps manifest signatures of multiple phase lines and result in H-T phase diagrams that are clearly anisotropic. Analysis of M(T) and M(H) data allow for the identification of the two lower temperature magnetically ordered states as antiferromagnetically ordered, with the moments aligned along the crystallographic c-axis, and the higher temperature, $T_2 < T < T_1$, state as having a finite ferromagnetic component. In addition, the metamagnetic transition at low temperatures, for H applied along the crystallographic c-axis (H$_{\parallel c}$) appears to be a near classic example of a spin-flop transition, resulting in a field stabilized antiferromagnetic state with the moments ordered perpendicular to the c-axis. Although the small moment ordering, and existence of multiple phase transitions in field and temperature, suggesting an energetic proximity of these states, could foretell a degree of pressure sensitivity, our measurements of R(T) for applied pressures up to 2.0 GPa indicate that there is very little pressure dependence of $T_1$, $T_2$ and $T_3$.

**Introduction**

Fe-based superconductors, cuprate based superconductors and Ce/U-based, Kondo-lattice based superconductors all have what is thought to be exotic, non-electron-phonon-mediated superconductivity located close to the suppression of magnetic, usually antiferromagnetic, order.[1] This observation has led to the idea that the suppression of fragile magnetic states may be a necessary (but not sufficient) requirement to discovering new families of superconducting materials. A fragile magnet is one that can have the ordering temperature as well as the size of the ordered moments suppressed by perturbation (i.e., doping, pressure, applied field, etc.).[1] Unfortunately, whereas most rare earth based intermetallic compounds tend to manifest antiferromagnetic (AFM) order, many of the transition metal based, metallic compounds with reduced ordered moments are ferromagnetic (FM). This is "unfortunate" because the avoided quantum criticality theoretically predicted and discussed over the past 20 years for metallic systems,[2-8] really does seem to be an experimental reality. For example, recent systems that we examined as part of our efforts to study and understand fragile magnets: $LaCrGe_3$,[9-11] $La_5Co_2Ge_3$[12] and even $YbFe_2Zn_{20}$[13] and $CeTiGe_3$[14] all have avoided FM quantum critical points (QCPs). To this end, we have been trying to identify or discover systems that start out as promising, small moment, transition metal based *antiferromagnets* with the intent to then use pressure and/or doping to perturb them.

Whereas $La_2Ni_7$ has been know structurally since at least 1969 [15] and studied for its magnetic properties for the past several decades [16-24], it has, so far, been only studied in polycrystalline form. As a result of this, the existing H-T phase diagrams are either an average or an admixture of the anisotropic H-T phase diagrams associated with the field applied along or perpendicular to the crystallographic c-axis. In 1983 Buschow [16] studied $La_2Ni_7$ as well as $La_2Ni_7H_x$; whereas the $La_2Ni_7$ was found to have a feature associated with AFM ordering below ~54 K in its temperature dependent, Curie-Weiss-like magnetic susceptibility, the $La_2Ni_7H_x$ sample had very small, essentially temperature independent, susceptibility data. As second study in 1983 by Parker and Oesterreicher[17] identified a $T_N$ of ~51 K and pointed out that the paramagnetic Weiss temperature of 70 K was more consistent with a FM than an AFM. In addition, Parker and Oesterreicher found that, "$La_2Ni_7$ exhibits the typical S-shaped magnetization versus field behavior of metamagnetic materials" [17] and were able to construct a H-T phase diagram of a single curve separating the paramagnetic from the AFM region. The paramagnetic to AFM phase line ran roughly from 60 kOe at base temperature to zero at $T_N$. Given the small, high field, saturated moment of ~0.1 $\mu_B$/Ni, La$_2$Ni$_7$ appeared to be an itinerant AFM system. Between 1993 and 2004 there was a series of papers focused on the M(T) and M(H) data,[19-23] and the inferred H-T phase diagram, associated with polycrystalline samples that were annealed at or near 900 °C for up to 5 weeks (so as to get as close to single phase, hexagonal La$_2$Ni$_7$ samples as possible). Initially, only a single phase line in the H-T phase diagram, very similar to that seen by Parker and Oesterreicher,[17] was detected. More detailed measurements (between 1997 and 2000) [20-22] revealed two, low field phase transitions: ~66 K and ~54 K and a H-T phase diagram with multiple regions was constructed.

One attempt was made in 1997 to microscopically detect AFM ordering via powder neutron diffraction. When the experiment failed to detect any signature of the AFM order upon cooling below $T_N$, the authors suggested an upper limit of 0.03 $\mu_B$/Ni [20], a value much smaller than the measured saturated moment of ~ 0.1 $\mu_B$/Ni. Very recently, [24] computational work has predicted a T = 0 K, modulated AFM structure with moments of less than 0.3 $\mu_B$ (but much larger than 0.03 $\mu_B$ given by [20]) pointing along the c-axis, forming FM-like blocks that alternate over a relatively long length scale.

Whereas M(T) and M(H) data have been studied by multiple groups, there is little or no other data in the literature. Temperature dependent electrical resistivity and thermoelectric power were measured on polycrystalline LaNi$_x$ compounds so as to provide 4f$^0$ (La) analogues to a study CeNi$_x$ materials. As part of this La$_2$Ni$_7$ was measured and data were presented in ref [18] without comment. There does appear to be a signature of a transition near 50 K in both electrical resistivity and thermoelectric power plots. To our knowledge there is no published specific heat as a function of temperature data. So as to start to better evaluate La$_2$Ni$_7$ both as a small, moment, itinerant magnetic system and also as a possible fragile magnetic system we have grown relatively large single crystals and used temperature dependent specific heat as well as temperature and field dependent magnetization and electrical resistivity to determine that there are three, zero field magnetic phase transitions at $T_1 = 61.0 \pm 0.2$ K, $T_2 = 56.5 \pm 0.2$ K and $T_3 = 42.2 \pm 0.2$ K, determine the anisotropic H-T phase diagrams for H$_{\parallel c}$ and H$_{\perp c}$, and evaluate the pressure dependence of the transition temperatures for hydrostatic pressures up to 2.0 GPa.

**Crystal growth and La-Ni phase diagram associated with La$_2$Ni$_7$ formation.**

Single crystals of La$_2$Ni$_7$ were grown out of a La-rich (relative to La$_2$Ni$_7$) binary, high-temperature melt. Elemental La (Ames Laboratory, 99.99+% pure) and Ni (Alpha, 99.9+% pure) were weighed out in a La$_{33}$Ni$_{67}$ atomic ratio and placed into a tantalum crucible which was sealed with solid caps on each end and a fritted cap in the middle to act as a frit or filter for decanting.[25,26] The assembled Ta crucible was then itself sealed into an amorphous silica tube with silica wool above and below it to act as cushioning. This growth ampoule was then place in a resistive box furnace. The furnace was then heated to 1150 °C over 10 hours, held at 1150 °C for 10 hours, cooled to 1020 °C over 4 hours and then very slowly cooled to 820 °C over 300 hours at which point the growth ampoule was removed and decanted in a centrifuge to separate the La$_2$Ni$_7$ single crystals from the residual liquid.[26] Crystals grew as well faceted plates with clear hexagonal morphology (see insets to fig. 1). Powder x-ray diffraction spectra (fig. 1) were taken at room temperature on a Rigaku Miniflex diffractometer with Cu-K$\alpha$ radiation are well fit to the hexagonal, Ce$_2$Ni$_7$, hP36, space group 194, structure.[27,28]

It should be noted that (i) the above growth is the result of some degree of optimization, and (ii) solution growth is sometimes the final arbiter of disagreements between composition-temperature phase diagrams, specifically the location of liquidus lines. In our initial attempts to grow La$_2$Ni$_7$ we used the most recent La-Ni binary phase diagram in the ASM-online data base [29,30] which has the exposed liquidus line for La$_2$Ni$_7$ existing between 63% at. Ni at 979 °C and 57% at. Ni and 802°C. When we cooled a melt of La$_{38}$Ni$_{62}$ from 1050 °C to 820 °C and decanted [26] we found that the melt was still in a single phase, liquid state. When we re-melted and the material, slow cooled and then decanted again at 750 °C we found a mixture of solid LaNi$_3$ and La$_7$Ni$_{16}$ (in roughly a 7:3 ratio as suggested by powder x-ray diffraction) in addition to decanted liquid. These results are inconsistent with refs. [30, 31] and suggest that the liquidus line for La$_2$Ni$_7$ is shifted to higher Ni concentrations. An earlier assessment of the La-Ni binary phase diagram [32] places the liquidus line for La$_2$Ni$_7$ between ~ 68% at. Ni and 976 °C and ~65% at. Ni and 811°C. When we performed growth using a starting composition of La$_{33}$Ni$_{67}$ and cooled from 1020 °C to 820 °C we produced single phase La$_2$Ni$_7$ single crystals allowing for an evaluation of the decanted liquid composition which was ~ 65 % at. Ni. It should be noted that the work reported by ref. [32], the lanthanum used was, "prepared at the Materials Preparation Center, Ames Laboratory, Iowa State University"; the same, very high purity La was used in for our crystal growth.

**Experimental methods**

Temperature dependent specific heat, Cp(T), anisotropic, temperature and magnetic field dependent measurements of electrical resistivity, $\rho$(T,H) and magnetization, M(T,H) were carried out in Quantum Design Physical Properties Measurement System (PPMS), and Magnetic Properties Measurement Systems (MPMS and MPMS3). Anisotropic d.c. magnetization data (M(T) in 1, 10 and 50 kOe as well as M(H) for several temperatures) were measured in MPMS classic systems and the cascades of M(T) and M(H) curves for multiple H and T values were measured in the MPMS3 using the VSM option. We normalized M(T,H) measured in the VSM option by the d.c. magnetization data to correct for potential differences in the relative accuracy of the VSM data.

Electrical resistivity was measured using a standard 4-probe geometry with contacts between the sample and Pt wire being made using Epotek-H20E silver epoxy. The samples were cut in long thin bars and the measurements were performed in a QD PPMS, on warming, with a rate of 0.25K/min and with a current excitation of 3 mA and frequency of 17 Hz. On average, the room temperature sample resistance was ~15m$\Omega$ and the contact resistance was ~ 3 $\Omega$. The current was applied in plane (perpendicular to the crystallographic c-axis) and perpendicular to magnetic field in both, $H_{\parallel c}$ and $H_{\perp c}$, configurations. Heat capacity measurements were made using semi-adiabatic thermal relaxation technique as implemented in the heat capacity option of the Quantum Design PPMS.

The temperature dependent resistivity of $La_2Ni_7$ was measured for applied hydrostatic pressures up to ~ 2 GPa. The measurements were made with the with current applied perpendicular to the c-axis direction in a Quantum Design physical property measurement system (PPMS) using a 3-mA excitation with frequency of 17Hz on cooling rate of 0.25K/min. A standard linear four-terminal configuration was used. The magnetic field was applied along the c-axis direction. To apply pressures up to ~ 2 GPa, a Be-Cu/Ni-Cr-Al hybrid piston-cylinder cell, similar to the one described in Ref. [33], was used. A 4:6 mixture of light mineral oil: n-pentane, which solidified at room temperature in the range of 3-4 GPa[33-35], was used as a pressure medium. Pressure values at low temperature were inferred from $T_c$(p) of elemental Pb[36, 37].

**Results**

Temperature dependent, thermodynamic and transport measurements on single crystalline La$_2$Ni$_7$ reveal signatures of three distinct phase transition temperature T$_1$ = 61.0 ± 0.2 K, T$_2$ = 56.5 ± 0.2 K and T$_3$ = 42.2 ± 0.2 K. Fig. 2 presents the specific heat divided by temperature, C$_p$(T)/T, as a function of temperature for 2 K < T < 130 K. Three phase transitions are resolvable, but quite small. The entropy under each feature (very roughly taken as the area between each anomaly and the extrapolated curve ignoring the anomalies) is less than 0.001(7Rln2); note that there are 7 Ni per formula unit. Although this is only a qualitative evaluation of entropy, and more entropy removal would be expected at temperatures below the ordering temperatures (i.e. magnons), this small entropy removal makes it very clear that, if these are magnetic transition, they will be associated with quite small ordered moments. A non-magnetically ordering, isostructural analogue would be needed for a more accurate evaluation of entropy changes and, unfortunately, one is not readily available. At low temperatures the specific heat follows a C(T) = γT + βT$^3$ temperature dependence for T$^2$ < 40 K$^2$ (upper inset) giving γ ∼ 40 mJ/mole-K$^2$ and β = 0.83 mJ/mole-K$^4$, which gives a Debye temperature of ∼ 280 K. Whereas the Sommerfeld coefficient, γ ∼ 40 mJ/mol-K$^2$, has a somewhat high value for a compound with 7 Ni or 9 atoms total per formula unit, i.e. 5.7 mJ/(mol-Ni-T$^2$) or 4.4 mJ/(mol-atomic-T$^2$), this value is being extracted well below the ordering temperatures, after the entropy removal associated with the magnetic ordering. With these caveats, some enhancement of the Sommerfeld coefficient, γ, was suggested by the recent band structure calculations in ref [24] which found an enhanced density of states (DOS) at the Fermi energy (E$_F$).

Fig. 3 presents the temperature dependent electrical resistivity, ρ(T) for 2 K < T < 300 K for two crystals with the current flowing perpendicular to the c-axis. The difference in inferred resistivity values are a measure of the geometric uncertainties in the distance between the voltage contacts and the cross sectional area. The upper inset shows an expanded view for 2 K < T < 70 K and the lower inset shows dρ(T)/dT plotted for 30 K < T < 70 K. Whereas there is a clear loss of spin disorder scattering seen in the ρ(T) data upon cooling through ∼ 65 K, the dρ(T)/dT data show three clear transition temperatures.[38] The residual resistivity ratio, RRR = ρ(300 K) / ρ(2 K), is greater than 18, indicating a relatively small amount of disorder scattering.

Fig. 4 presents the anisotropic, H = 1 kOe, temperature dependent magnetization divided by applied field, M(T)/H, as well as the polycrystalline average, (M/H)$_{poly}$ = (1/3) M/H$_{\|c}$ + (2/3)

M/H$_{\perp c}$. The inset to Fig. 4 shows an expanded view for 35 K < T < 75 K. Whereas for M/H$_{\|c}$ three transition temperatures are readily seen, for M/H$_{\perp c}$ the signatures of the phase transitions are more subtle, especially the lowest, ~ 42 K, one. (M/H)$_{poly}$ can be fit to a Curie Weiss, C/(T+θ) + χ$_0$, temperature dependence, with a temperature independent, χ$_0$, for 80 K < T < 300 K. From this fit we find θ = -55 K, χ$_0$ = 1.0 x 10$^{-3}$ emu/mol-Ni, and from C = 0.20 emu-K/mol-Ni, we get μ$_{eff}$ = 1.3 μ$_B$/Ni, a value comparable, but somewhat larger than found in earlier polycrystalline measurements.[17]

In order to determine the values of the three transition temperatures, in figs. 5 and 6 we compare the C(T), dρ(T)/dT,[38] and d[(M(T)/H)T)]/dT [39] data in the vicinity of the transitions; whereas dρ(T)/dT, and d[(M(T)/H)T)]/dT are related to Cp(T) in the vicinity of a paramagnetic to antiferromagnetic phase transition, they can be helpful in identifying phase transition temperatures for cascades of magnetic transitions.[40-42] Fig. 5 shows the features associated with the lowest transition, T$_3$, with a clear peak seen in all three of the data sets. The transition temperature is inferred from the position of the local maximum, giving a value of T$_3$ = 42.2 ± 0.2 K. In fig. 6 the signatures of the upper two transitions are shown. For C$_p$(T)/T and dρ(T)/dT there are well resolved peaks with maxima located at T$_1$ = 61.0 ± 0.2 K, T$_2$ = 56.5 ± 0.2 K. Whereas for T$_3$ there is fair agreement between the value of the transition temperature inferred from C$_p$(T)/T and dρ(T)/dT and the transition temperature inferred from d[(M(T)/H)T)]/dT , for T$_2$ and T$_1$ the features in d[(M(T)/H)T)]/dT, especially for the H∥c data, are somewhat shifted. As will be discussed below, the magnetization data for H∥c indicates that between T$_1$ and T$_2$ the low field state has a ferromagnetic component, making the use of d[(M(T)/H)T)]/dT to determine a transition into or out of this state a little less accurate. In total, then, based on these data, La$_2$Ni$_7$ has three transitions upon cooling in zero (or low) field: T$_1$ = 61.0 ± 0.2 K, T$_2$ = 56.5 ± 0.2 K, and T$_3$ = 42.2 ± 0.2 K.

**Magnetic field - temperature phase diagrams**

All prior H-T phase diagram work has been based on polycrystalline samples that have had either only one or, at most, two low field transition temperatures identified. Given that H-T phase diagrams of systems that have multiple, potentially complex and/or fragile magnetic phases, are often anisotropic, the use of single crystalline samples is strongly preferred. In fig. 7a and 7b we present the H –T phase diagram for $H_{\parallel c}$ and $H_{\perp c}$. These phase diagrams were inferred from the R(H) and M(H) isothermal sweeps as well as R(T) and M(T) constant field sweeps shown in figs. 8 – 17 below. From these data we have been able to identify and track multiple transitions with resolvable and intelligible features. There are also smaller features or more subtle transitions that we are not completely identifying; some of these are discussed and commented on in the text below. The aim of the current paper is to lay out the primary features of these anisotropic phase diagrams. Clearly further work and study will be needed to fully delineate and understand the $La_2Ni_7$ system. Fig. 8a presents the $M(H_{\parallel c})$ data for increasing field. There is an intelligible complexity to the primary features associated with these curves; three primary phase lines can be seen evolving in the M(H) data. The transition to the saturated paramagnetic state is located near 63 kOe at base temperature and moves monotonically downward with increasing temperature until no longer being resolvable near $T_1$. The second, low temperature transition, near 30 kOe, is relatively temperature insensitive for 2 K < T < 20 K, but then splits into two transition features with one decreasing to H = 0 near $T_2$ the other decreasing to H = 0 near $T_3$. All of these features are quite visible in the M(H) data and transition fields are identified via maxima in analysis of dM/dH plots (not shown). In addition to the three primary lines mentioned above, there is an apparent separation of the line that extrapolates to $T_2$ as it drops from 20 kOe toward H = 0. Although this line would seem to extrapolate to H = 0 near 50 K, there are no signatures of a transition at this temperature in either $C_p(T)$, $\rho(T)$ or lower field M(T)/H data. This suggests that there may be a missing, low field dome in this region, but it is not readily resolved or systematically followed.

Fig. 8b presents an expanded view of the higher temperature, lower field, $M(H_{\parallel c})$ data. If we start with the 50 K data near 20 kOe (upper right corner of the figure) we can see the transition to paramagnetic / saturated paramagnetic state into the C-phase; near 6 kOe there is a transition from the C-phase into the lower field B state. As temperature increases to 52.5 K and then 55 K both of these transition fields decrease. At 55 K there is still a small, low field region of M(H) data with a lower slope that can be associated with the B-phase, followed by the step like rise in M(H) associated with the transition into the C-phase. All of these data consistently suggest that the C-phase has a finite, net ferromagnetic component to its ordered state. The M(H) data for 57.5 K and 60 K have a low field saturation that is consistent with being in the C-phase from lowest measured field. Indeed, this is consistent with the low field M(T) data shown in fig. 4 above as well as fig. 12 below.

Fig. 9 presents the $\rho(H_{||c})$ data; given that there is significant temperature dependence of the zero field $\rho(T)$, the $\rho(H)$ isotherms separate from each other rather naturally (fig. 9a). At lowest temperatures (fig. 9d) there are two clear transitions visible in the $\rho(H)$ data, one between 30-35 kOe and the other between 60 – 65 kOe. As temperature rises both of these transition fields decrease. For T = 40 K, fig. 9c, three phase transition features are seen and for higher temperatures, fig. 9b, two features are resolved up to 55 K; at 60 K only one feature is seen and for 65 K no features in $\rho(H)$ are resolved. As can be seen in fig. 7a, the transition fields inferred from the M(H) and $\rho(H)$ data agree with each other very well.

Fig. 10 presents the $M(H_{\perp c})$ isotherm data for increasing field. Having already understood the $M(H_{||c})$ data in fig. 8, and tracking three phase lines as they go to H = 0 as T increases, we can follow a similar strategy for the $M(H_{\perp c})$ data. Given that the critical fields needed to induce metamagnetic phase transitions shift to higher values for $H_{\perp c}$, the transition to the saturated paramagnetic state only comes into our 70 kOe range for T = 22.5 K. As temperature is increased to higher values, this highest field, metamagnetic phase transition is induced at lower and lower values of applied field, reaching H = 0 near $T_1$. The T = 22.5 K data also show a lower field transition near ~ 60 kOe. This feature moves up in field for lower temperatures, just barely manifesting below 70 kOe for 12.5 K. As temperature increases above 22.5 K, the lower field transition moves down in field and separates into two, broad features, clearly seen, for example in the 37.5 K isotherm. These two features head toward H = 0 at different rates with one phase line extrapolating toward $T_2$ and the other toward $T_3$. As was the case for the data shown in fig. 8, all of these features are quite visible in the M(H) data and transition fields are identified via maxima in analysis of dM/dH plots (not shown). In addition to these more conspicuous features there appear to be a pair of phase lines running from the lower-field $T_2$ or $T_3$ lines up to the $T_1$ line for intermediate fields and temperatures.

Fig. 11 presents the $R(H_{\perp c})$ isotherm data; again, the temperature dependence of the zero field resistivity data leads to a natural off-set between the isotherms. From T = 2 K to T = 20 K (fig. 11d) there are two distinct features, starting near 80 and 75 kOe at base temperature and decreasing to ~72 and ~62 kOe by 20 K. It should be noted that the lower of these two transitions manifests clear, field up / field down hysteresis that grows smaller with increasing temperature (similar, but smaller hysteresis can also be seen in Fig. 9). For T = 25 K to T = 45 K (fig. 11c) there are three transitions observable, in some cases with the middle transition only clearly revealing itself by comparing field up and field down curves and letting the hysteresis highlight the subtle feature. In fig. 11b there are two features visible for T = 50 and 55 K and a single feature visible for T = 60 K. As can be seen in fig. 7b, the transition fields inferred from the M(H) and ρ(H) data agree with each other very well.

Whereas the M(H) and ρ(H) isotherm data tend to be more sensitive to H-T phase lines that are more horizontal in nature, and therefore, often, offer greater detail for the lower temperature parts of the phase diagram, the constant magnetic field M(T) and ρ(T) data tend to be more sensitive to phase lines that are more vertical in nature and therefore, often, offer greater detail for the higher temperature parts of the phase diagram. In fig. 12 the M(T)/H sweeps at constant $H_{\|c}$ are shown for fields ranging from 1 to 65 kOe. The (H,T) data points we extract from these measurements, via identification of extrema in d[(M(T)/H)T)]/dT plots (not shown), agree well with the M(H) and ρ(H) data point already appearing in fig. 7a. Starting at highest fields and low temperatures we can see the $T_1$ line as a sharp transition from the ordered state into the paramagnetic state (or saturating paramagnetic state) move from ~ 12 K at 60 kOe up to just under $T_1$ at 1 kOe. For applied fields of 30 kOe and below the $T_3$ and $T_2$ lines and the associated features in the M(T) data become apparent. A pair of lower temperature steps in the M(T) data move upward in temperature as the applied field decreases, reaching just below $T_3$ and $T_2$ for $H_{\|c}$ = 1 kOe. There are finer features in the M(T) data shown in fig. 12 that we show in fig. 7a, such as the slight splitting of the $T_1$ line in the 25-40 K region. These may delineate very narrow regions of other phases or may be artifacts that we do not yet understand. For this first determination of the anisotropic, H-T phase diagrams we will focus on the more conspicuous and less ambiguous features in our data. As will be discussed below, there may well be further work needed to fully understand the interplay between all the phases that may exist in $La_2Ni_7$. As discussed above, with regards to fig. 8b, the C-phase appears to have a well-defined ferromagnetic component to its ordering. This is particularly apparent in the lowest field M(T) data for $T_2 < T < T_1$.

In fig. 13 we present $\rho(T)$ data taken for differing $H_{\|c}$ values and in fig. 14 we present the $d\rho(T)/dT$ data. Given that (i) the $\rho(T)$ data change a lot over the 2 K < T < 70 K temperature range and (ii) the effects of magnetic ordering as well as applied field are resolvable, but small compared to the temperature dependence, it is difficult to see the systematic effects of applied field in the bare $\rho(T)$ data. In fig. 14 the $d\rho(T)/dT$ data reveal a systematic shift of transitions with applied field and, using the local maxima to identify critical temperatures, we can see very good agreement with the other (H,T) data on fig. 7a. We can again see the three, primary $T_1$, $T_2$, and $T_3$ lines as well as some of the finer structure we found in our other measurements. It is worth noting that the $\rho(T)$ data for $H_{\|c}$ = 60 and 65 kOe does not clearly reveal the $T_1$ phase line; this is not unusual, given that, as mentioned earlier, $\rho(T)$ data better reveals the more vertical lines on an H-T phase diagram with the $\rho(H)$ data more clearly revealing the more horizontal ones.

Similar data and analysis can be collected and analyzed for the $H_{\perp c}$ direction. Fig. 15 presents $M(T)/H$ data for constant $H_{\perp c}$. Starting at our highest applied field, 70 kOe, there are two clear transitions visible in the M(T) data that, from $d[(M(T)/H)T)]/dT$ plots, can have transitions temperatures of 11 K and 21 K identified. The higher temperature feature increases in temperature as $H_{\perp c}$ is decreased, ending up at ~ $T_1$ for $H_{\perp c}$ = 1 kOe. The lower temperature feature also moves up in temperature as $H_{\perp c}$ is decreased, reaching $T_3$ for $H_{\perp c}$ = 1 kOe. For fields between 70 and 55 kOe two distinct features can be seen in the M(T) data; below 55 kOe a third and sometimes a somewhat less distinct fourth or fifth feature can be seen. Below 30 kOe three dominant, well defined features separate and become clear with the middle on ending near $T_2$ for $H_{\perp c}$ = 1 kOe. These data are plotted on fig. 7b and agree well with the data extracted from the M(H) and $\rho(H)$ sweeps. As has been mentioned before, there are some finer structures in the M(T)/H data for constant $H_{\perp c}$ that we are not currently quantifying; these may, at some future date reveal further structures.

In fig. 16 we present $\rho(T)$ data taken for differing $H_{\perp c}$ values and in fig. 17 we present the $d\rho(T)/dT$ data. These data reveal well defined features that allow for the identification of transition temperatures. At highest fields, a single transition becomes detectable for H = 65 kOe and moves up in temperatures as the field is decreased to 45 kOe. The extracted transition temperatures match well with the $T_3$ line. It is worth noting that $\rho(T)$ does not seem to be sensitive to the higher temperature features that were detected by our other measurements, again illustrating the need to use multiple types of measurements and sweeps to fully determine a H-T phase diagram. As field is lowered below 45 kOe two and then ultimately three distinct features emerge, ultimately clearly separating into the $T_3$, $T_2$ and $T_1$ lines. Fig. 7a and 7b show the general good agreement between all of the data points determined from the $\rho(H)$, $M(H)$, $\rho(T)$, $M(T)/H$ data sets.

**Pressure dependence of transition temperatures.**

In order to make an initial assessment of the pressure sensitivity of $La_2Ni_7$, we measured the temperature dependent electrical resistance for applied pressures, p < 2 GPa, in a self-clamping, piston-cylinder cell. In fig. 18 we show R(T) for 2 K < T < 300 K with the upper inset showing and expanded range centered on 42 K and the lower inset showing an expanded range centered on 60 K. These data immediately reveal that the three phase transitions, as well as the temperature dependent resistance data as a whole, are not very sensitive to pressures up to 2 GPa. Fig. 19 presents the dR(T)/dT plots for 20 K < T < 80 K for the data shown in fig. 18. For most pressures we can resolve features associated with the three phase transitions. In fig. 20 we plot the pressure dependence of the three magnetic phase transition temperatures; indeed, as was already suggested by fig. 18 and 19, there is very little change in the transition temperatures with pressure. Given that the position and sharpness of these features can change with applied field, in fig. 21a, we plot R(T) for 30 K < T < 65 K with a field of 10 kOe applied along the c-axis; in fig. 21b we plot the dR(T)/dT of the same data. The 10 kOe transition temperature data are also plotted in fig. 20; it can be seen that, (i) the field dependence of the ambient pressure data is consistent with the T-H phase diagram for $H_{\|c}$ shown in fig. 7a (i.e. the $T_2$ line being much more sensitive to 10 kOe than either the $T_1$ or $T_3$ line) and (ii) that there is very little change in the transition temperatures with pressure at either 0 or 10 kOe.

**Discussion**

The growth of large, single crystalline samples of $La_2Ni_7$ has allowed for zero field measurements of $\rho(T)$ and $C_p(T)$ combined with low field measurements of $M(T)/H$ to identify three zero (or low) field, magnetic phase transition temperatures, $T_1 = 61.0 \pm 0.2$ K, $T_2 = 56.5 \pm 0.2$ K and $T_3 = 42.2 \pm 0.2$ K. Detailed, anisotropic M(T,H) and $\rho$(T,H) measurements have allowed for the construction of anisotropic H-T phase diagrams, revealing multiple regions, labeled A-F in fig. 7a and 7b. Whereas phases A and B appear to be AFM in nature, phase C clearly has some ferromagnetic component, most likely combined with some finite-q ordering vector. The low-temperature, saturated moment of ~ 0.12 $\mu_B$/Ni, as well as the very small change in entropy associated with the features in specific heat suggest small moment, itinerant moment ordering. This is further supported if we follow the Rhodes-Wohlfarth [43] analysis outlined in ref [22]. The parameter is $\mu_c/\mu_{sat}$ where $\mu_{eff}^2 = \mu_c(\mu_c + 2\mu_B)$; if we use our $\mu_{eff} = 1.3$ $\mu_B$/Ni and our $\mu_{sat} = 0.12$ $\mu_B$/Ni we find $\mu_c = 0.64$ $\mu_B$/Ni and $\mu_c/\mu_{sat} = 5.3$, with is consistent with an itinerant system.

Although the experimental reality of La$_2$Ni$_7$ is much more complex than a single transition to an AFM ground state, it is useful to compare our results to recent bandstructural work. In their study of La$_2$Ni$_7$ and Y$_2$Ni$_7$, Crivello and Paul-Boncour, used electronic bandstructure calculations to gain insight into their magnetically ordered states.[24] For both compounds they found that a ferromagnetic state was the most stable, low temperature state, with a lowering of total energy by 5 meV/Ni for each compound. In addition, the ordered moments were found to favor alignment along the crystallographic c-axis. This is consistent with a FM transition of ~ 50 K in Y$_2$Ni$_7$, but clearly is not consistent with the AFM ground-state found for La$_2$Ni$_7$. For La$_2$Ni$_7$ there was a nearby (energetically) antiferromagnetic state with blocks of Ni moments aligned parallel and antiparallel to the c-axis. The energy difference between this AFM state and the aforementioned FM state is less than 1 meV/Ni and considered to be within the accuracy of the DFT calculations. As such, Crivello and Paul-Boncour [24] claim that both magnetic structures present the same stability at 0 K. Given that the FM blocks that make up the computationally predicted AFM state consist of 6 layers of Ni atoms that are ferromagnetically aligned, our measured, paramagnetic theta of -55 K (as opposed to a positive value for a simple AFM) is not too disconcerting.[24] Of course, the fact that we have determined that the highest temperature, C-phase, has a clear FM component to its ordering provides an experimental rational for the sign of theta as well.

Although the computational work only examines a single magnetically ordered state, these band structure results are consistent with several aspects of our data. Given the computational degeneracy between AFM and FM states for La$_2$Ni$_7$, the multiple, zero field transitions, as well as field induced transitions are not surprising. Indeed, the C-phase shown in fig. 7a and 7b has a clear, net ferromagnetic component that is replaced, at lower temperatures by the AFM states in the B-and A-phases. In addition, the measured, linear component of the temperature dependent specific heat, Cp/T, $\gamma$ = 40 mJ/mol K$^2$, is somewhat enhanced and is consistent with the computationally predicted enhanced DOS at $E_F$. Finally, our anisotropic M(T) and M(H) data demonstrate that the ordered moments (in low fields) are aligned along the c-axis. As will be discussed below, the low field alignment of the moments along the crystallographic c-axis is even further demonstrated a clear spin-flop transition that is associated with the D-phase.

The pressure insensitivity of all three magnetic phase transitions is rather surprising for such a small moment, antiferromagnetic ordering where, naively, some degree of fragility might be anticipated. These results suggest that La$_2$Ni$_7$ may be rather incompressible and, indeed, DFT based band structure calculations, using the conducted using the PBE exchange-correlation functional,[44] suggest (qualitatively) that La$_2$Ni$_7$ has a significantly larger bulk modulus than either EuCd$_2$As$_2$ or LaCrGe$_3$,[45] two recently studied compounds with well-defined pressure dependences. [9-11, 46] Whereas our results for applied pressures up to 2 GPa clearly suggest the need for higher pressure measurements, our current data also suggest that this may be increasingly difficult given the subtlety of the features in resistivity, the smallness of the features in specific heat and difficulty of measuring and tracking antiferromagnetic phase transitions with magnetization measurements for pressures above 2 GPa. Empirically, a comparison of La$_2$Ni$_7$ and Y$_2$Ni$_7$ offers mixed signals. The crystal structures are similar and there is some contraction of the volume per formula unit (consistent with a degree of positive chemical pressure). The transition temperatures of the two compounds are similar (implying a perhaps weak pressure dependence in ordering temperature) but the saturated moment size in Y$_2$Ni$_7$ is roughly a factor of two smaller than in La$_2$Ni$_7$ (both experimentally and computationally).[24, 47] Further theoretical / computational insight may be possible if neutron scattering measurements can determine the ordering wave vector associated with each of the zero field regions. Application of pressure in silica (computationally) may provide some insight as to what higher pressures will do, especially if the same simulations can capture the current ambient pressure magnetic structures and their pressure dependences.

Having constructed the phase diagrams shown in fig. 7a and 7b, we can see that there are clear similarities and differences. At the grossest level the two, H-T phase diagrams can be understood by observing that there have to be three, zero-field transition temperatures, with (H,T) lines emerging from them at low fields and, at high fields, as we approach T = 0 K, there are two critical fields with (H,T) lines emerging from them. At intermediate temperatures and fields, a more complex geometry of (H,T) lines emerge. There are two clear (H,T) lines: one runs from $T_1$ to what is labeled as $H_1$, the highest metamagnetic field at base temperature; the other runs from $T_3$ to the lower metamagnetic field at base temperature (and labeled as $H_3$). As we go from $H_{\|c}$ to $H_{\perp c}$ both $H_1$ and $H_3$ increase, with $H_3$ increasing by a much larger percentage, i.e. drawing closer to $H_1$. Whereas for $H_{\|c}$ there are four rather well defined H-T regions (with the open question of what is the nature of the apparent line between the low field $T_1$ and $T_2$ lines that itself does not reach down to H = 0), for $H_{\perp c}$ there appear to be five, with a skinny, lenticular region marked as F existing between the $T_1$ and $T_3$ lines at intermediate fields and temperatures.

Whereas the three, regions that extend down to H = 0 (A, B, and C in both fig. 7a and 7b) are associated with the same ordered states in the two phase diagrams, at least at lowest fields, the regions that exist only at finite fields (D, E, and F) are not inherently related. This said, we can compare the M(H) data for $H_{\|c}$ and $H_{\perp c}$ from fig. 8 and 10. Fig. 22 shows data for T = 15 K; for fields below 30 kOe, we can see a clear anisotropy in the M(H) with $H_{\perp c}$ having a significantly larger slope. This is consistent with the computational prediction that the low temperature, AFM ordered phase has the Ni moments aligned along the c-axis direction. For higher fields applied along the c-axis, the data shown in fig. 22 are a classic example of a spin-flop transition. [48,49] As such these data strongly suggest that for $H_{\|c}$, the D-phase has a similar arrangement of ordered moments as does the A-phase but with their orientation rotated by ~ 90 degrees. In this scenario, then, the ordered moments in the D-phase maintain AFM ordering but are aligned so as to be roughly perpendicular to the c-axis. In fig. 22, then, for 32 kOe <~ H <~ 56 kOe, for $H_{\|c}$ (D-phase) and for $H_{\perp c}$ (A-phase), the M(H) data are associated with an AFM order that is aligned roughly perpendicular to the respective applied field.

Whereas the D-phase appears to be AFM in nature, the metamagnetic-E-phase for $H_{\perp c}$ (seen in fig. 22 for $H_{\perp c}$ ~> 64 kOe as well as fig. 10 for other temperatures) likely has a ferromagnetic component to the order, i.e. the magnetic unit cell has a finite magnetization value. Clearly neutron scattering will be needed to determine the wave vectors associated with each of the phases identified in the phase diagrams delineated in fig. 7.

The data shown in fig. 22, in addition to providing some insight into the nature of the D-phase, can also be used to extract some initial estimates of the exchange field, $H_E$, and uniaxial anisotropy field, $H_A$, within the basis of the two sublattice Neel model. [48,49] Given that $La_2Ni_7$ is clearly a small moment, itinerant system with potentially complex order, i.e. having more than two sublattices, this analysis may be questionable, but it can, at least, provide some context. If we use the formalism presented by Holmes et al. [48], take the spin-flop field to be 31 kOe, take the saturated moment at high fields to be 0.12 $\mu_B$/Ni, and take the anisotropic susceptibilities to be the two slopes of the nearly linear M(H) data shown in fig. 22 for fields below 30 kOe, (giving $\chi_\parallel$ = 1.72 x $10^{-4}$ $\mu_B$/Ni-kOe and $\chi_{\perp c}$ = 9.38 x $10^{-4}$ $\mu_B$/Ni-kOe) and assume that $H_A$ is small compared to $H_E$ (as was done in [48]), we can infer that $H_E$ is 125 kOe and $H_A$ is 3 kOe. These values may provide some benchmark for future computational or neutron scattering efforts to better model or understand the finer details of the $La_2Ni_7$ structures.

In Summary, we have determined three, zero field, magnetic transition temperatures for $La_2Ni_7$: $T_1$ = 61.0 ± 0.2 K, $T_2$ = 56.5 ± 0.2 K and $T_3$ = 42.2 ± 0.2 K. These magnetically ordered phases are associated with small moments (~ 0.12 $\mu_B$/Ni in the saturated state) and small changes in entropy. Remarkably, $T_1$, $T_2$ and $T_3$ are relatively pressure insensitive (i.e. changing by less than 3 K) for applied pressures up to 2 GPa. We have determined anisotropic H-T phase diagrams for $H_{\parallel c}$ and $H_{\perp c}$. We have identified the ground state phase, A, as being antiferromagnetic with the moments aligned along the c-axis. For H∥c, as the applied field is increased to above ~ 33 kOe, the metamagnetic D-phase appears to be a spin-flop state with the ordered moments still antiferromagnetically aligned, but now perpendicular to the applied field. Whereas the B-phase appears to be antiferromagnetically ordered, the highest temperature, low field, C-phase has a clear ferromagnetic component. The E- and F-phases also have net ferromagnetic components.

Given the wealth of detail that our single crystal measurements have provided, multiple follow up measurements and experiments are suggested. Whereas neutron scattering measurements were tried on polycrystalline materials [20] and failed to detect any new wave vectors associated with the onset of AFM order, clearly new measurements on single crystalline samples are needed. With the information from our current work, as well as (potentially) ordering wave vectors from scattering measurements, bandstructural calculations should be revisited and revised using the new details outlined in this work to refine modeling of the magnetism in $La_2Ni_7$. In addition to these efforts, temperature dependent NMR as well as angle resolved photoemission spectroscopy (ARPES) measurements are possible. Both NMR and ARPES can also shed light onto the nature of the magnetic order and how it impacts the bandstructure.

Acknowledgements: We thank Linlin Wang for providing relative values of the compressibility $La_2Ni_7$, $LaCrGe_3$ and $EuCd_2As_2$.[45] This work was supported by the U.S. Department of Energy, Office of Basic Energy Science, Division of Materials Sciences and Engineering. The research was performed at the Ames Laboratory. Ames Laboratory is operated for the U.S. Department of Energy by Iowa State University under Contract No. DE-AC02-07CH11358. D.H.R. was supported in part by U.S. Department of Energy, Office of Basic Energy Science, Division of Materials Sciences and Engineering under Contract No. DE-AC02-07CH11358; financial support for D.H.R. was also provided by Fonds Quebecois de la Recherche sur la Nature et les Technologies, and the Natural Sciences and Engineering Research Council (NSERC) Canada. R.A.R. was supported in part by the Gordon and Betty Moore Foundation's EPiQS Initiative through Grant GBMF4411.

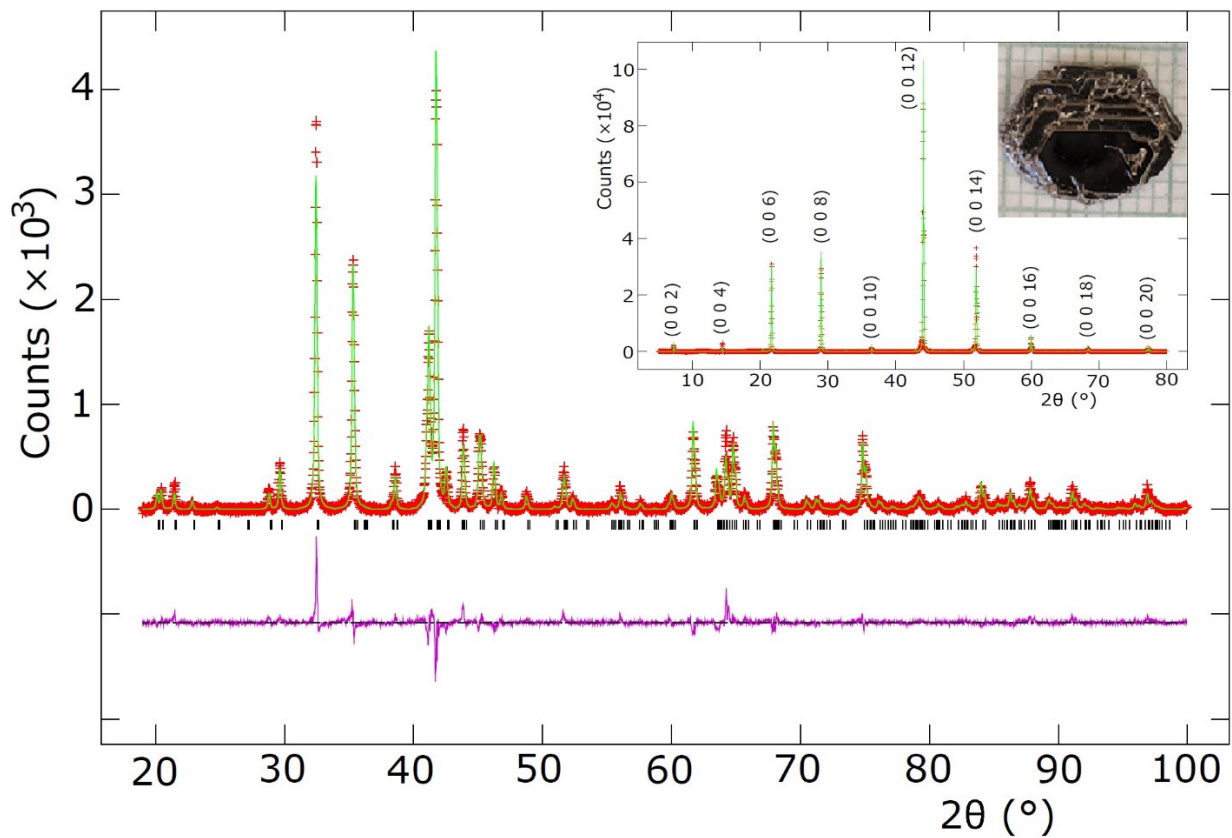

Fig. 1: Powder X-ray pattern for ground, single crystal $La_2Ni_7$. Using a Rietveld refinement,[27, 28] the lattice parameters of a = 5.06352(11) Å and c = 24.6908(8) Å were inferred. Inset shows data from a Bragg-Brentano diffraction from a single crystal plate demonstrating that the c-axis is perpendicular to the plate. The LeBail fit to the single crystal run gives c = 24.6991(3) Å. In both cases the error bars result from the fitting programs used. The image is a single crystal shown over mm-grid graph paper.

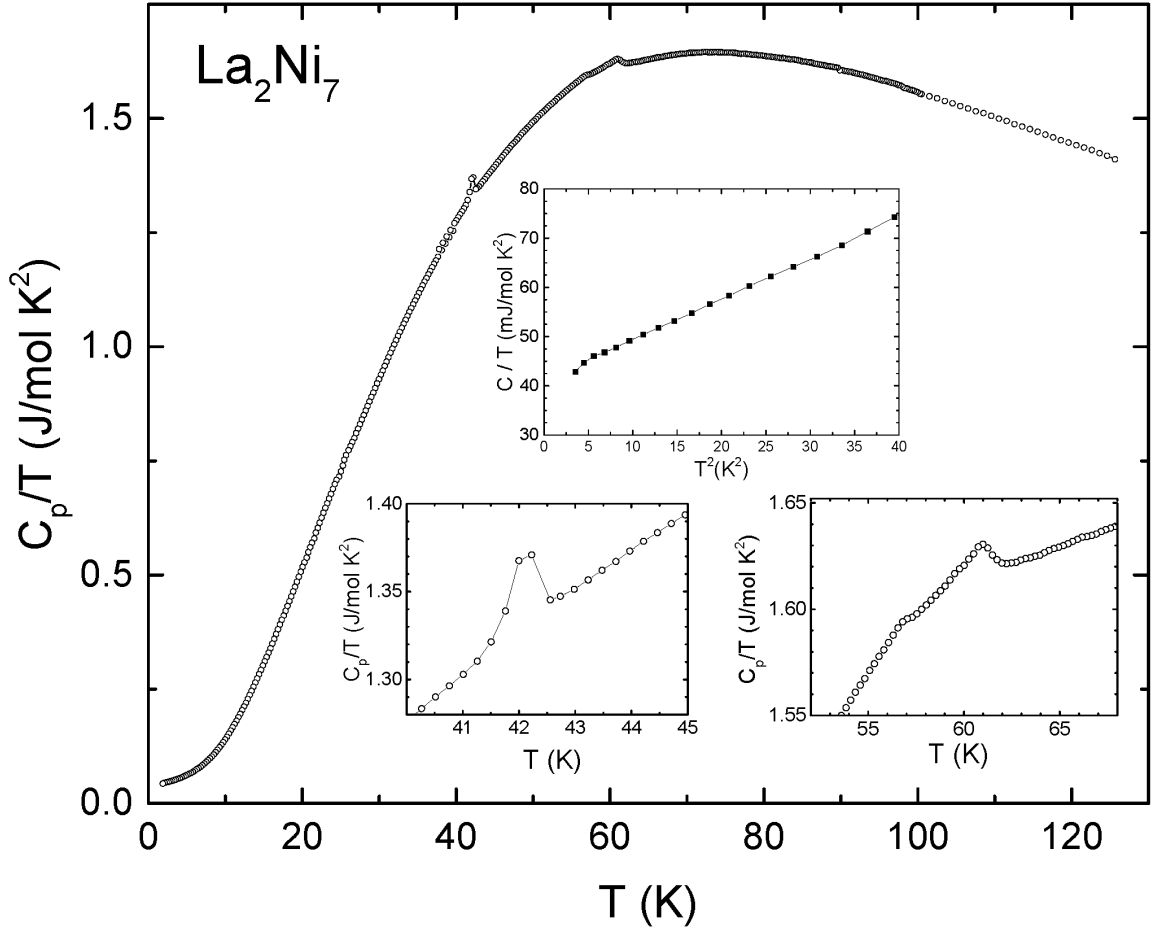

Fig. 2: Temperature dependent specific heat divided by temperature, $C_p/T$, of La$_2$Ni$_7$ for 2 K < T < 125 K. Upper inset, plot of $C_p/T$ versus $T^2$ for $T^2$ < 40 K$^2$. Lower left inset, expanded view for 40 K < T < 45 K; lower right inset, expanded view for 50 K < T < 70 K.

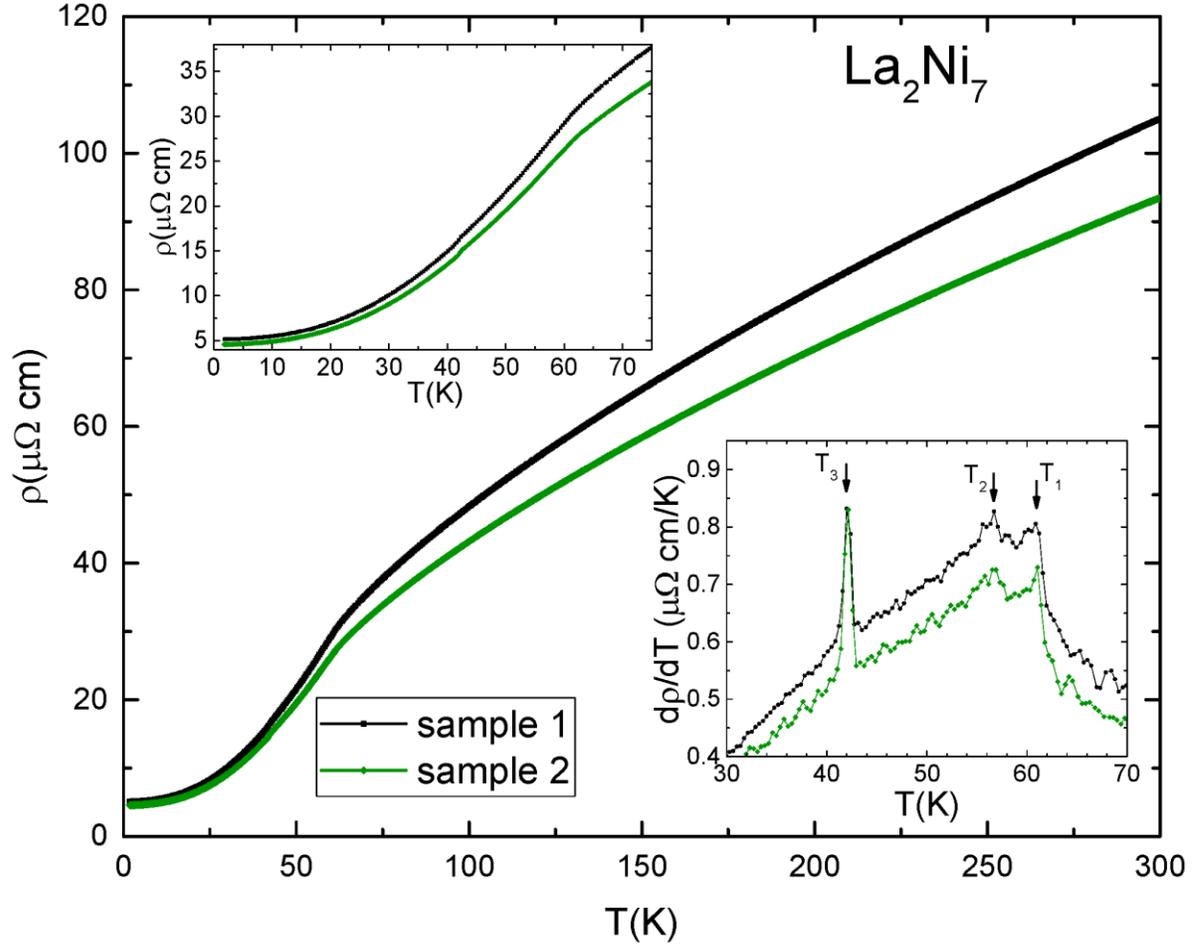

Fig. 3: Temperature dependent electrical resistivity of two samples of La$_2$Ni$_7$ for 2 K < T < 300 K. The different values of resistivity are representative of our geometric uncertainties in length between voltage contacts as well as the cross-sectional area of the samples. Upper inset, expanded view for 2 K < T < 70 K; lower inset: d$\rho$(T)/dT plotted for 30 K < T < 70 K with transition temperatures $T_1$, $T_2$, $T_3$ indicated by arrows.

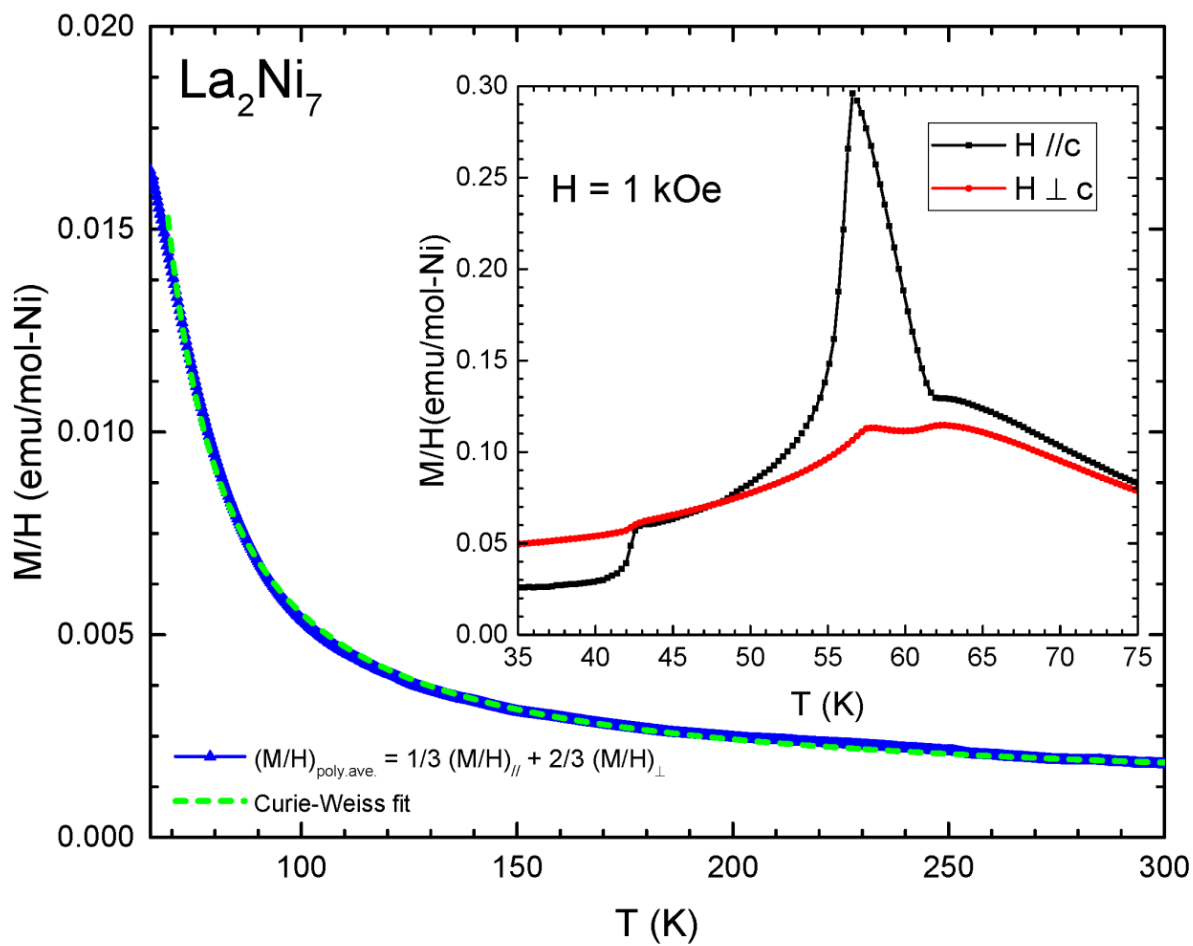

Fig. 4: Temperature dependent, anisotropic, low field magnetization divided by applied field, M(T)/H for 2 K < T < 300 K for field applied along the c-axis, $M/H_{\parallel c}$, for the field applied perpendicular to the c-axis, $M/H_{\perp c}$, and for $(M/H)_{poly} = 1/3\ (M/H_{\parallel c}) + 2/3\ (M/H_{\perp c})$. The green dashed line is a Curie-Weiss fit to the $(M/H)_{poly}$ data (see text). Inset shows and expanded view of 35 K < T < 75 K.

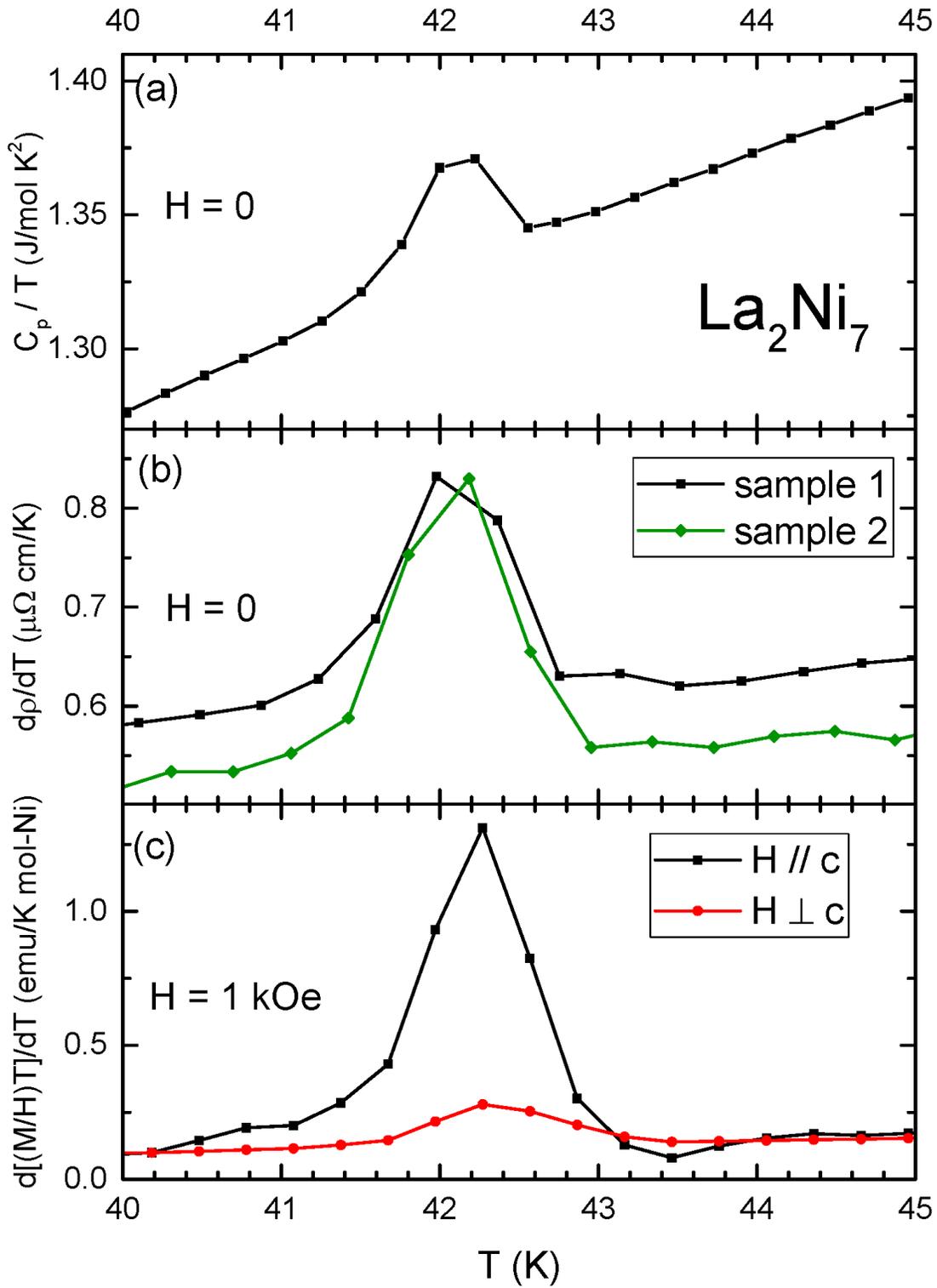

Fig. 5: $C_p/T$, $d\rho/dT$, $d[(MT)/H]dT$ data for 40 K < T < 45 K.

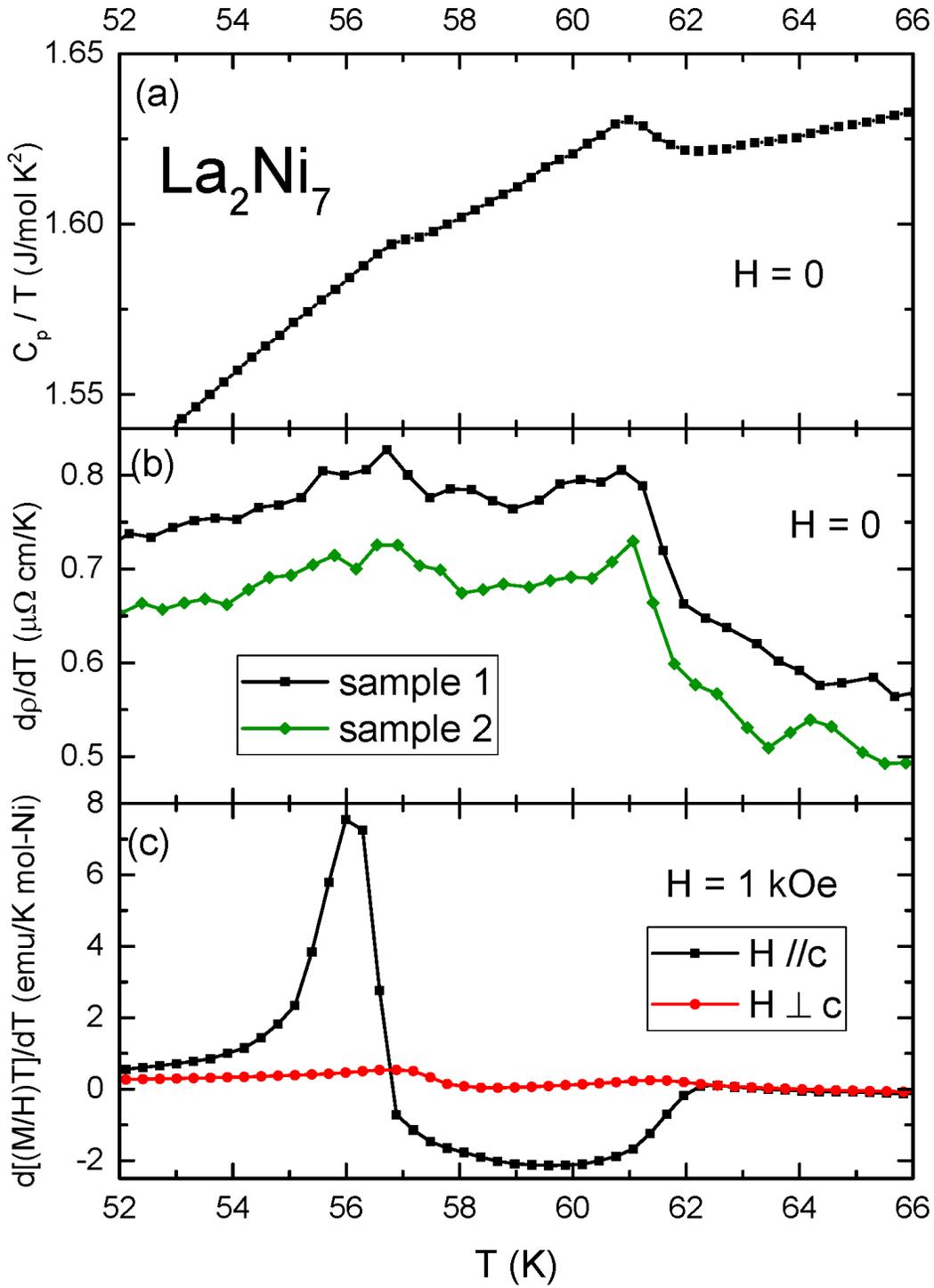

Fig. 6: $C_p/T$, $d\rho/dT$, $d(MT)/HdT$ data for 52 K < T < 66 K.

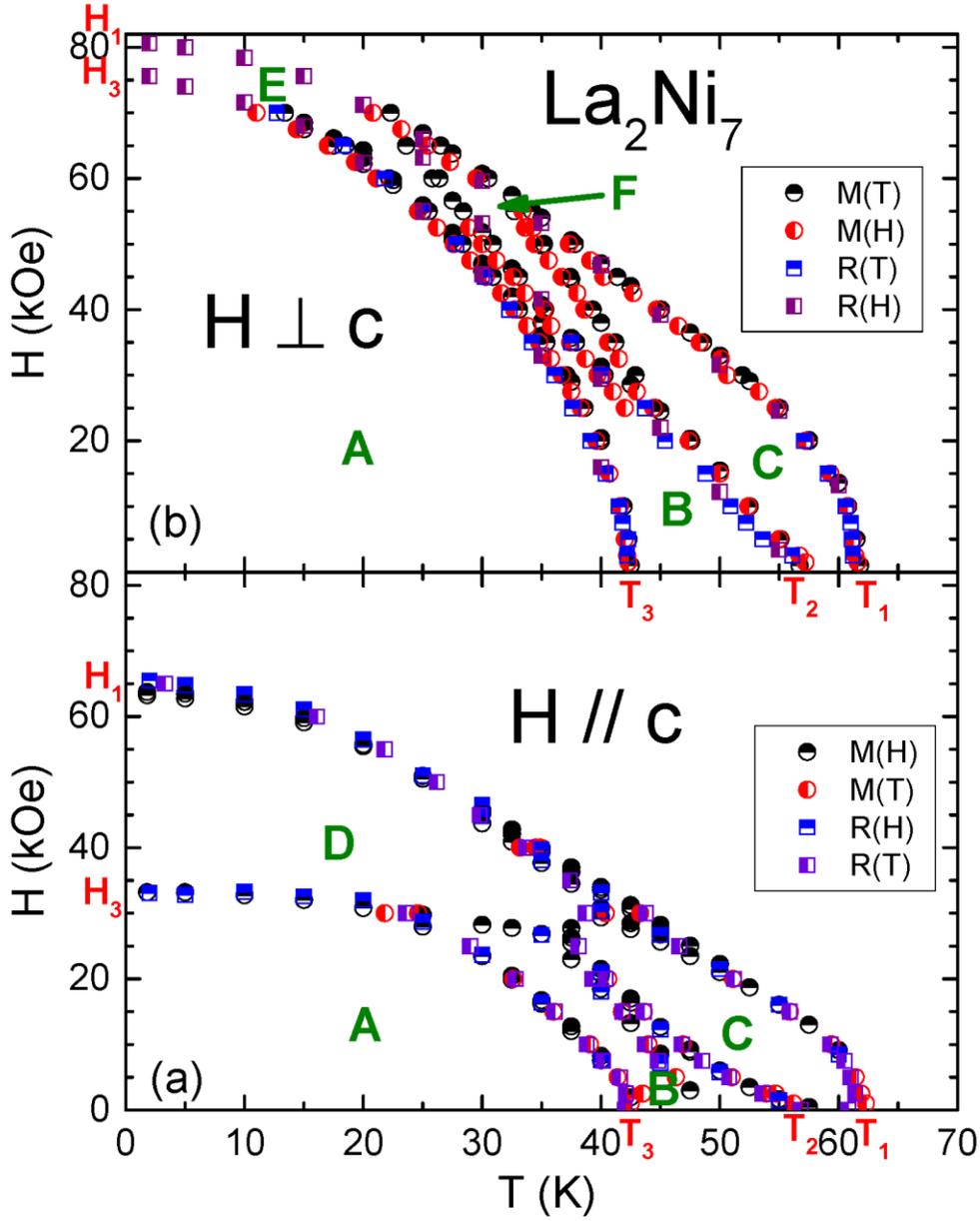

Fig. 7: H-T phase diagrams for $La_2Ni_7$ (a) for $H_{\parallel c}$, (b) for $H_{\perp c}$ constructed from M(T), M(H), $\rho(T)$ and $\rho(H)$ data with T or H increasing. Zero-field transition temperatures: $T_1 = 61.0 \pm 0.2$ K, $T_2 = 56.5 \pm 0.2$ K, and $T_3 = 42.2 \pm 0.2$ K are shown along the horizontal axis, low temperature, anisotropic, metamagnetic fields $H_1$ and $H_3$ are shown along the vertical axis. Whereas the A, B, and C phases have to be the same in the lowest field limit, the phases D, E, and F exist only at finite fields and need to be examined individually, see text for details.

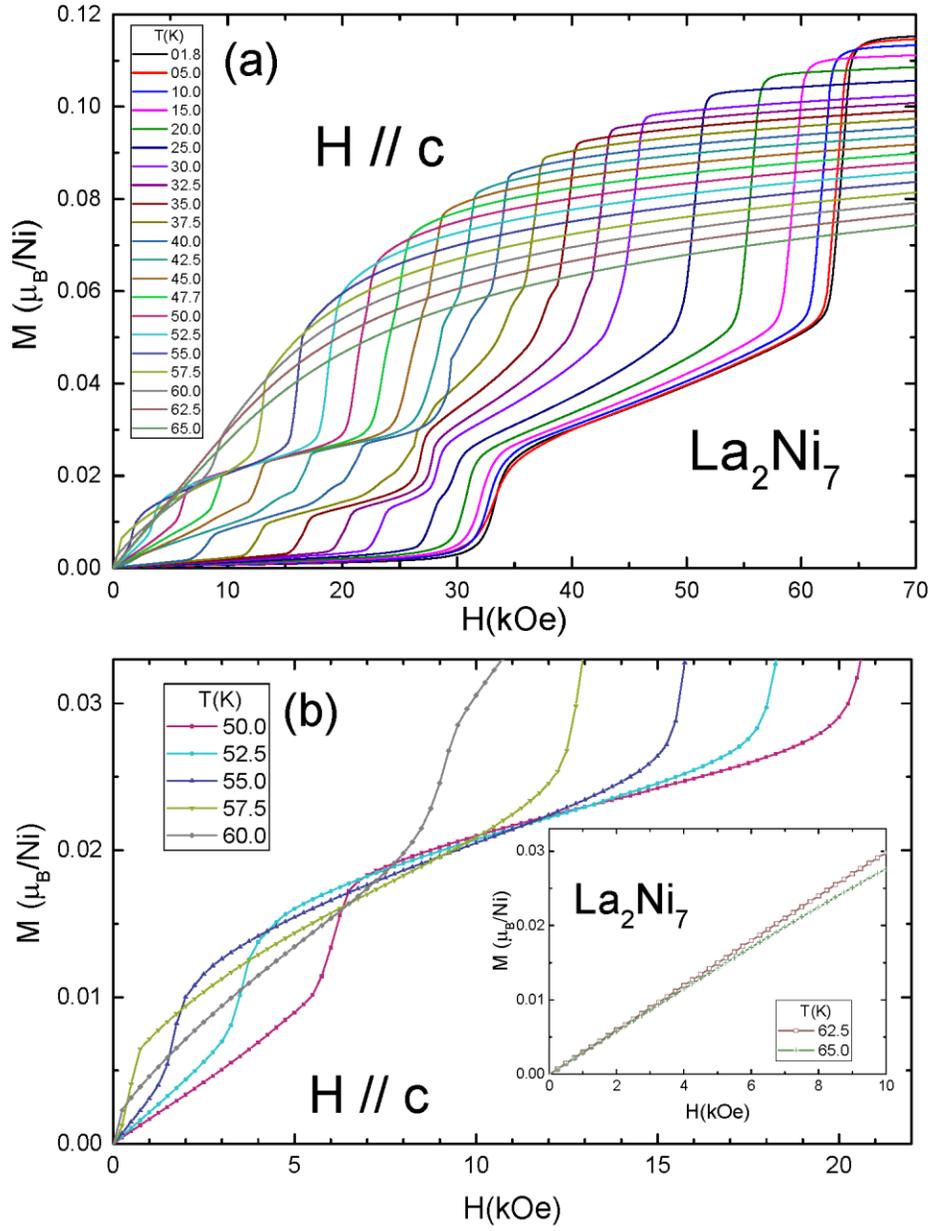

Fig. 8: (a) M(H) isotherms for $H_{\|c} < 70$ kOe (field increasing) and for selected T in the 1.8 K $\leq$ T $\leq$ 65 K range; (b) M(H) isotherms for $H_{\|c} < 22$ kOe and for selected T in the 50 K $\leq$ T $\leq$ 60 K range; inset: M(H) isotherms for $H_{\|c} < 10$ kOe and for selected T = 62.5 K and 65 K.

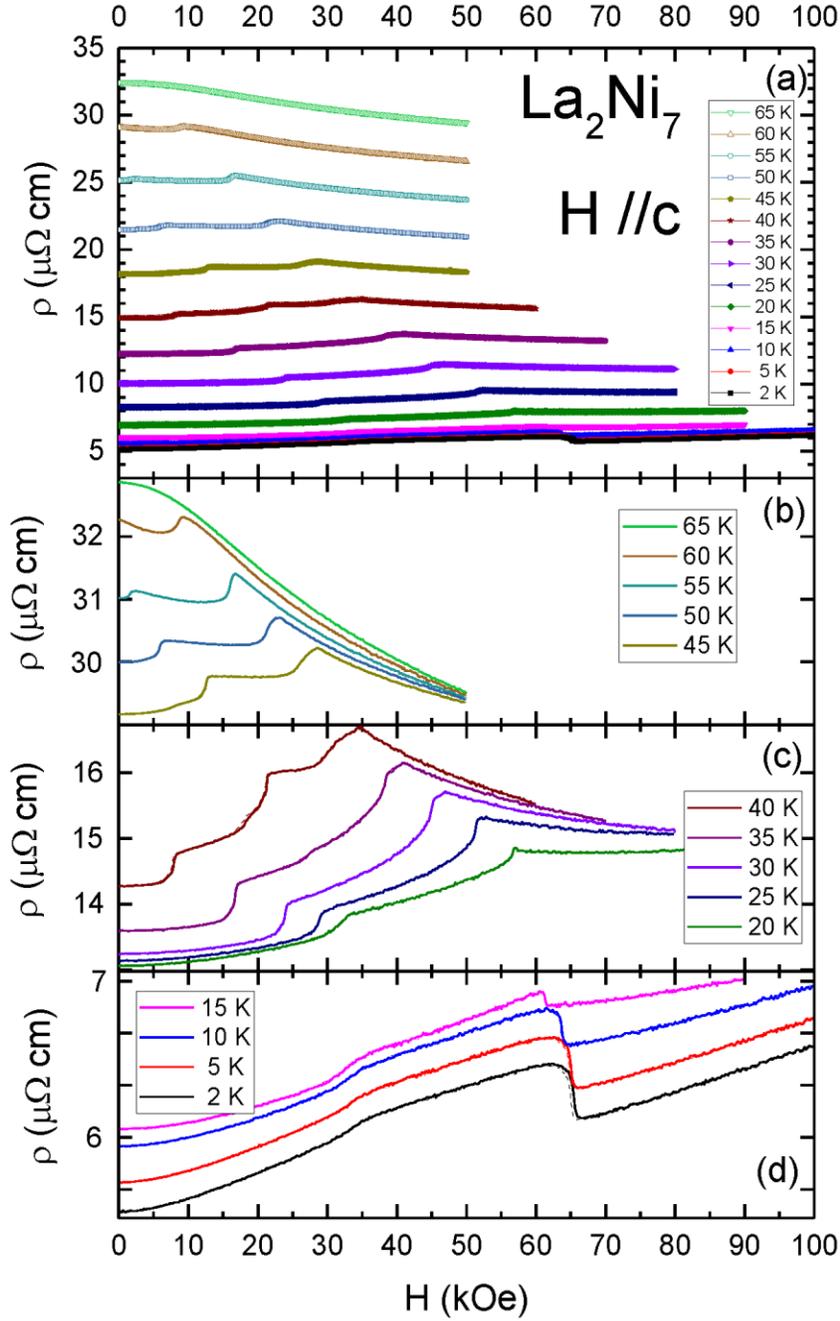

Fig. 9: (a) $\rho(H)$ isotherms for $H_{\|c} < 100$ kOe and for selected T in the 2 K <= T <= 65 K range; (b) $\rho(H)$ isotherms for $H_{\|c} < 50$ kOe and for T = 45, 50, 55, 60, 65 K with T = 45, 50, 55, 60 K data shifted along y-axis for clarity; (c) $\rho(H)$ isotherms for $H_{\|c} < 80$ kOe and for T = 20, 25, 30, 35, 40 K with T = 20, 25, 30, 35 K data shifted along y-axis for clarity; (d) $\rho(H)$ isotherms for $H_{\|c} < 100$ kOe and for T = 2, 5, 10, 15 K with T = 2, 5, 10 K data shifted along y-axis for clarity. Data taken on decreasing field are shown with dashed lines.

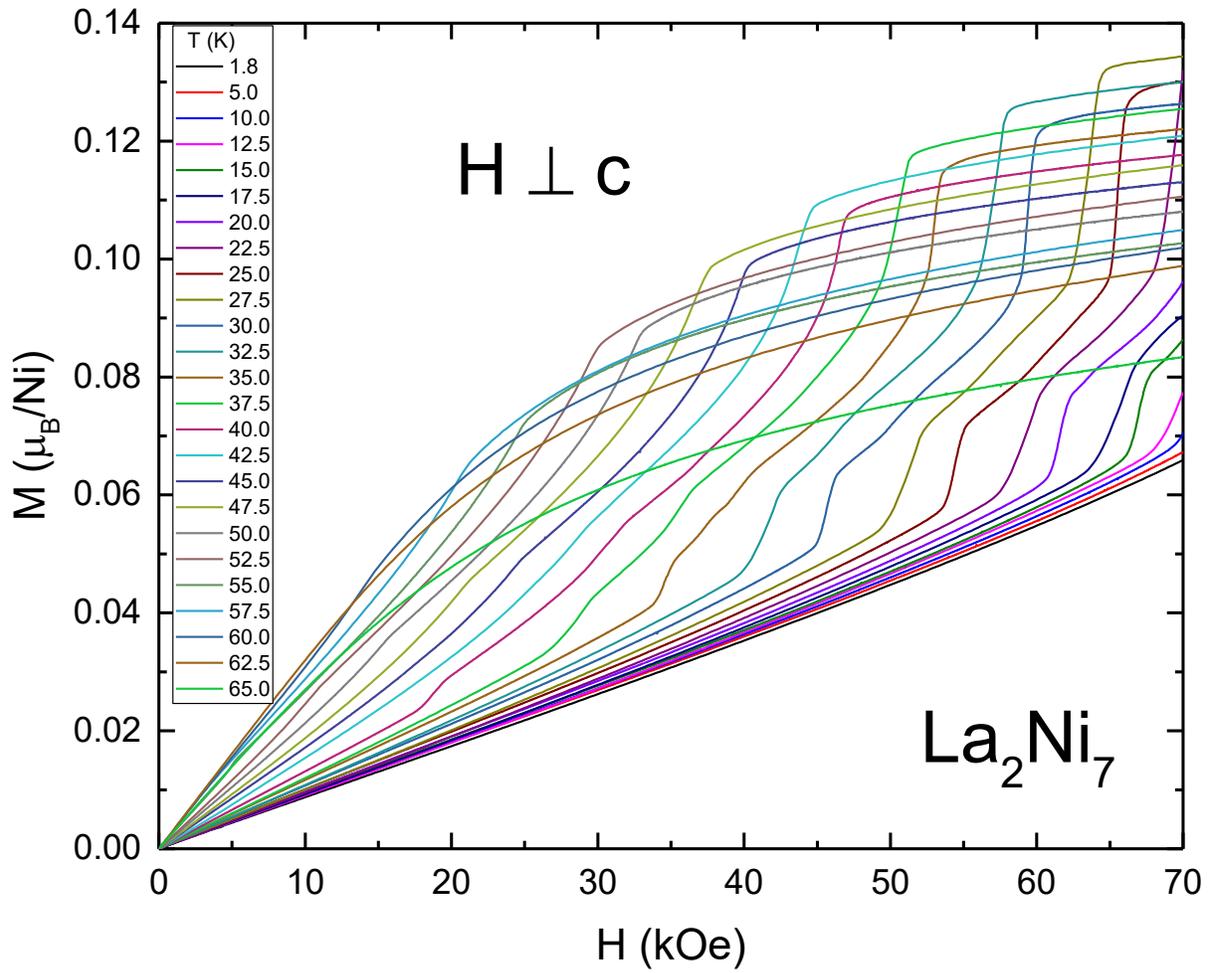

Fig. 10: M(H) isotherms for $H_{\perp c} < 70$ kOe (field increasing) and for selected T in the 1.8 K <= T <= 65 K range.

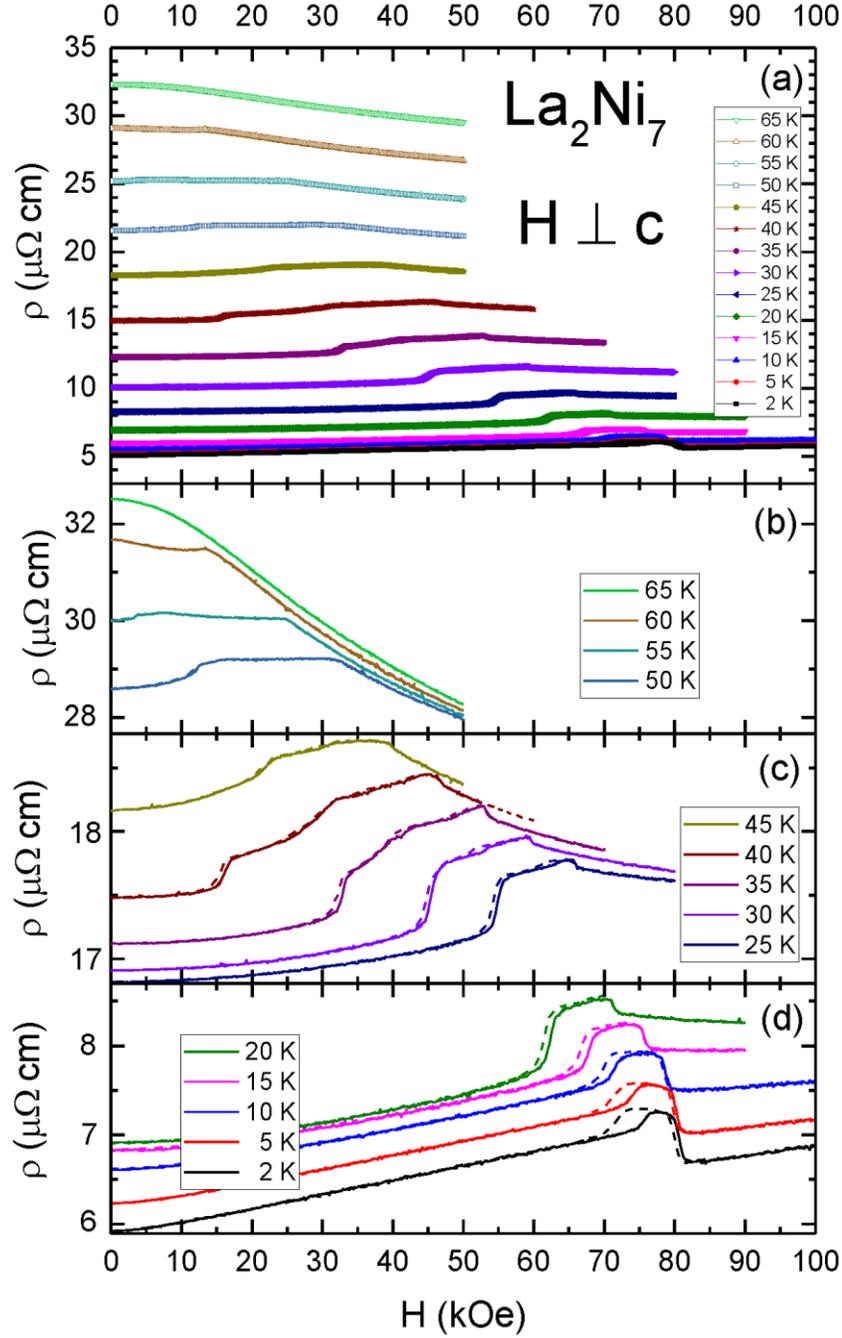

Fig. 11: (a) $\rho(H)$ isotherms for $H_{\perp c} < 100$ kOe and for selected T in the 2 K $\leq$ T $\leq$ 65 K range; (b) $\rho(H)$ isotherms for $H_{\perp c} < 50$ kOe and for T = 50, 55, 60, 65 K with T = 50, 55, 60 K data shifted along y-axis for clarity; (c) $\rho(H)$ isotherms for $H_{\perp c} < 80$ kOe and for T = 25, 30, 35, 40, 45 K with T = 25, 30, 35, 40 K data shifted along y-axis for clarity; (d) $\rho(H)$ isotherms for $H_{\perp c} < 100$ kOe and for T = 2, 5, 10, 15, 20 K with T = 2, 5, 10, 15 K data shifted along y-axis for clarity. Data taken on decreasing field are shown with dashed lines.

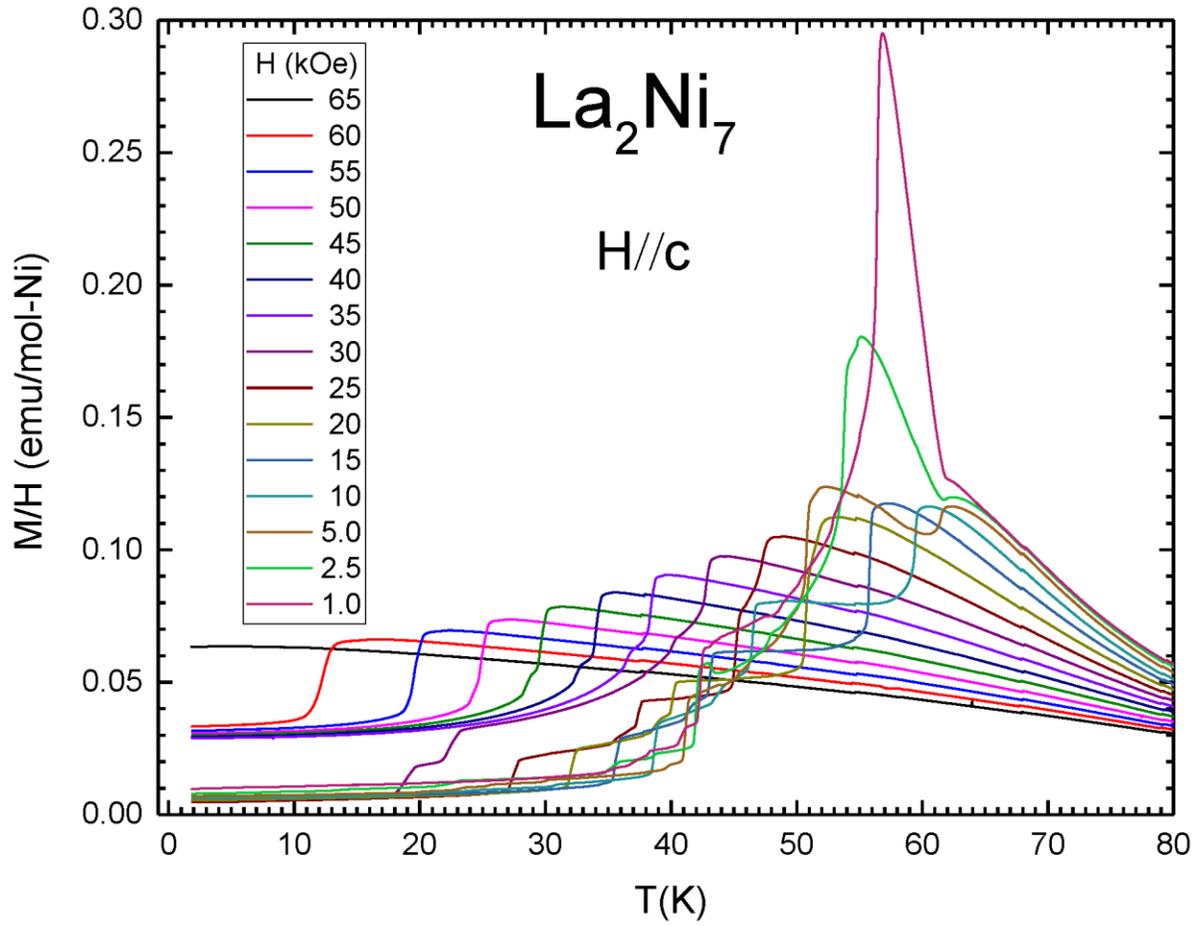

Fig. 12: M(T)/ $H_{\|c}$ data for 1.8 K ≤ T ≤ 80 K (temperature increasing) and selected fields 1 kOe ≤ $H_{\|c}$ ≤ 65 kOe.

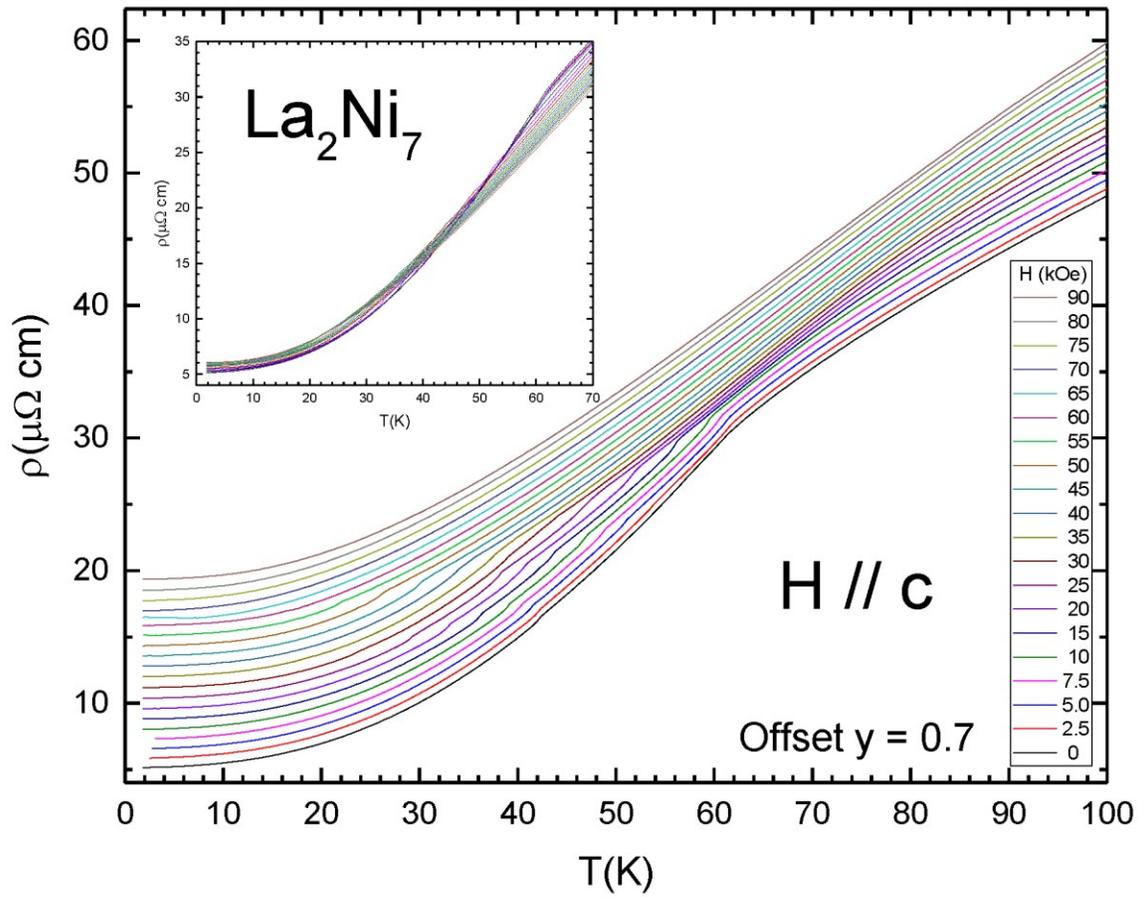

Fig. 13: $\rho(T, H_{\|c})$ data for 2 K ≤ T ≤ 100 K (temperature increasing) and selected fields 0 kOe ≤ $H_{\|c}$ ≤ 90 kOe. Data curves in the main figure are offset from each other by 0.7 μΩ-cm for clarity whereas the curves in the inset are not offset.

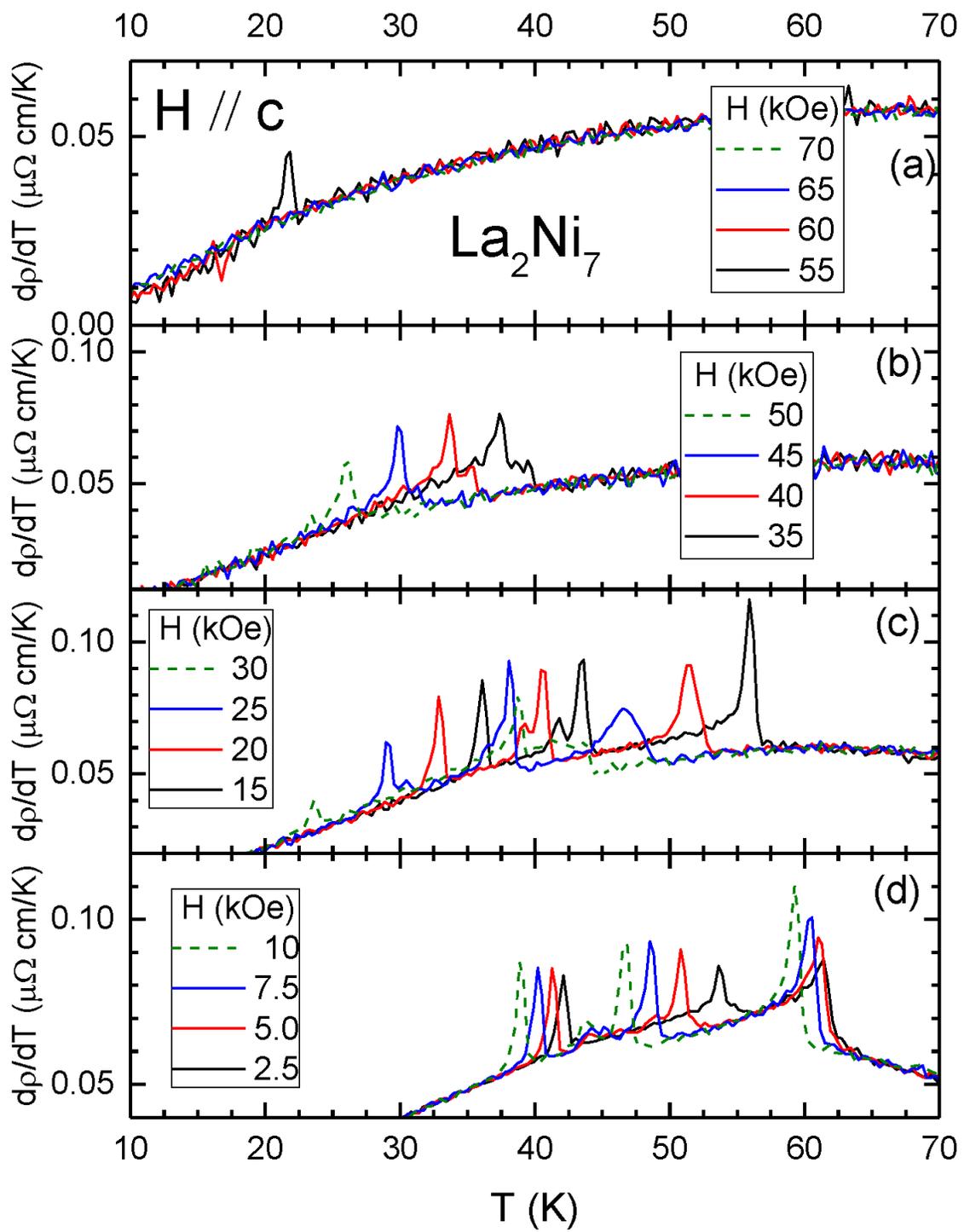

Fig. 14: dρ/dT plots for data shown in fig. 13.

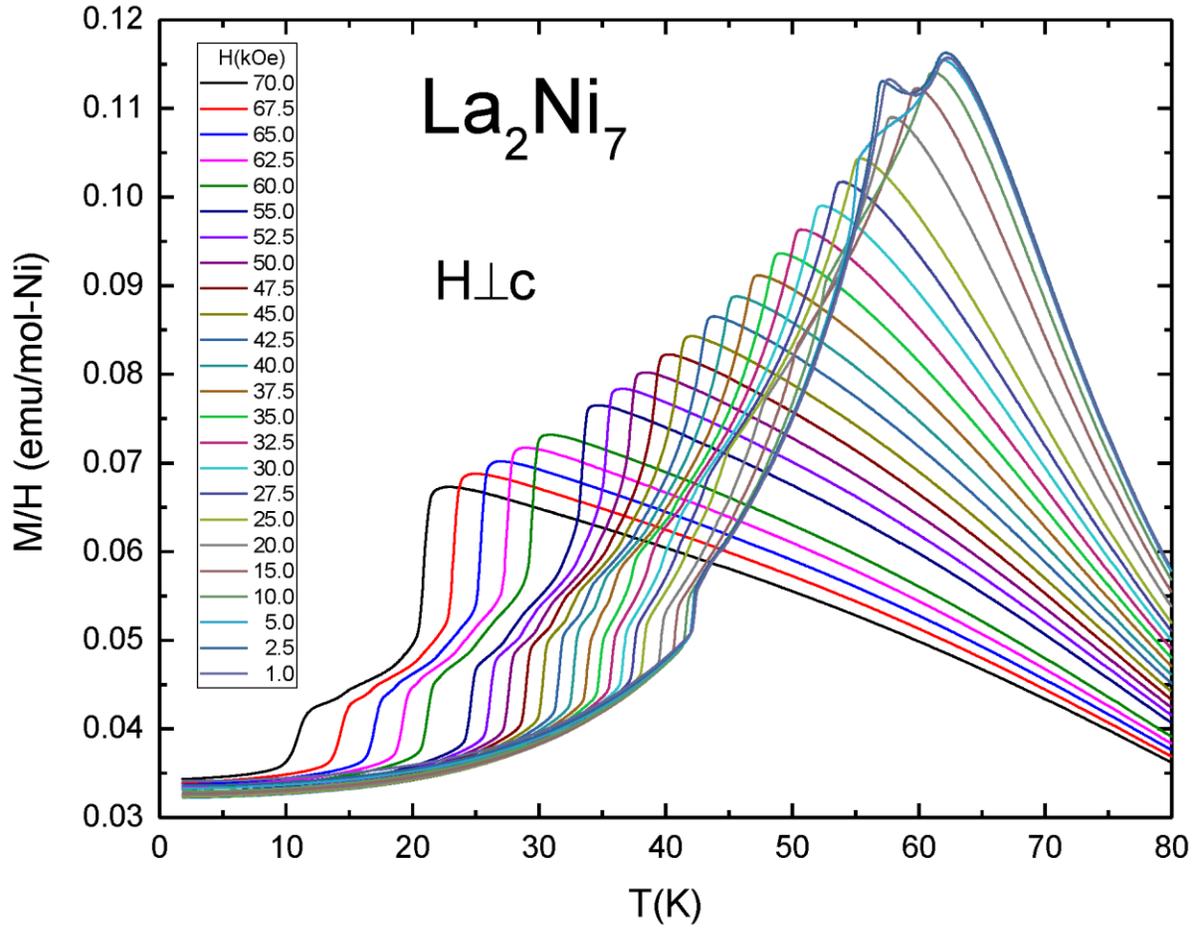

Fig. 15: M(T)/ $H_{\perp c}$ data for 1.8 K ≤ T ≤ 80 K (temperature increasing) and selected fields 1 kOe ≤ $H_{\parallel c}$ ≤ 70 kOe.

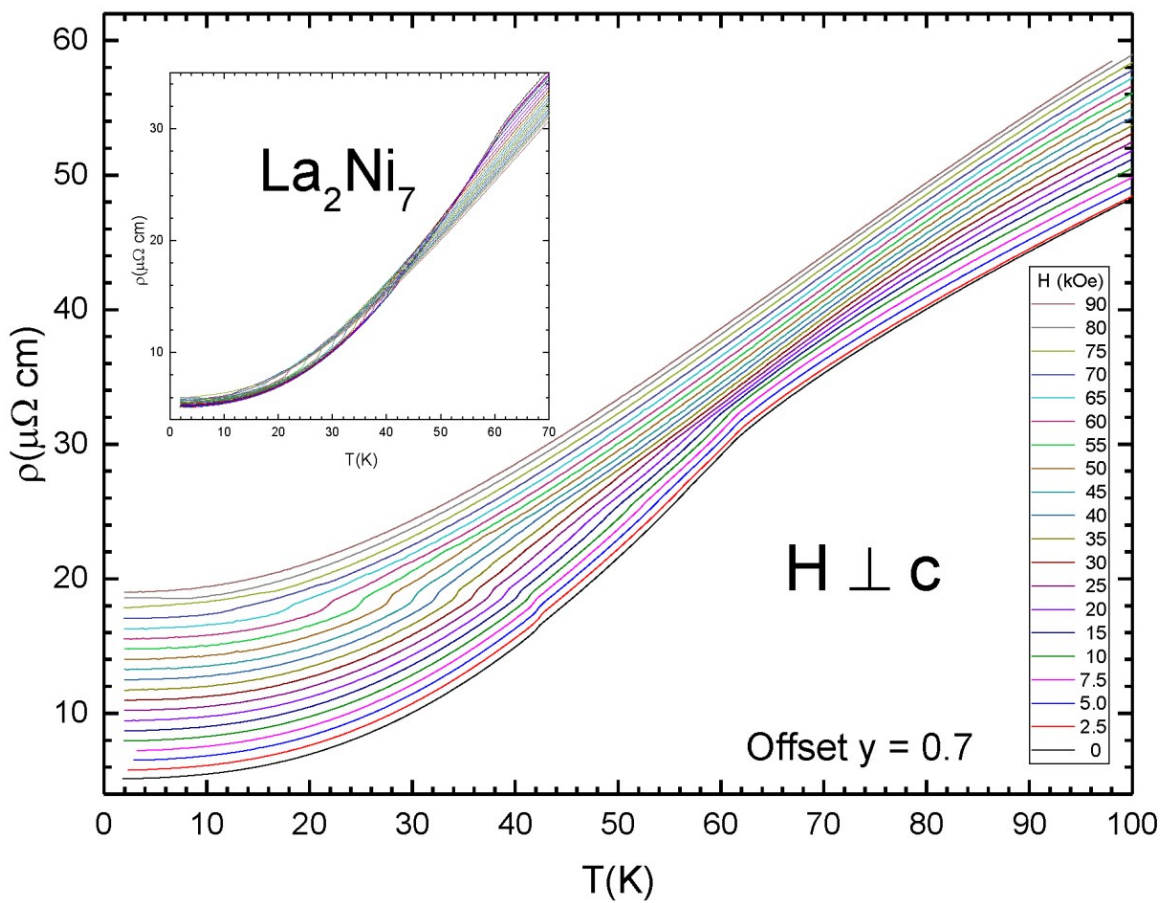

Fig. 16: $\rho(T)/ H_{\perp c}$ data for 1.8 K ≤ T ≤ 100 K (temperature increasing) and selected fields 0 kOe ≤ $H_{\|c}$ ≤ 90 kOe. Data curves in the main figure are offset from each other by 0.7 μΩ-cm for clarity whereas the curves in the inset are not offset.

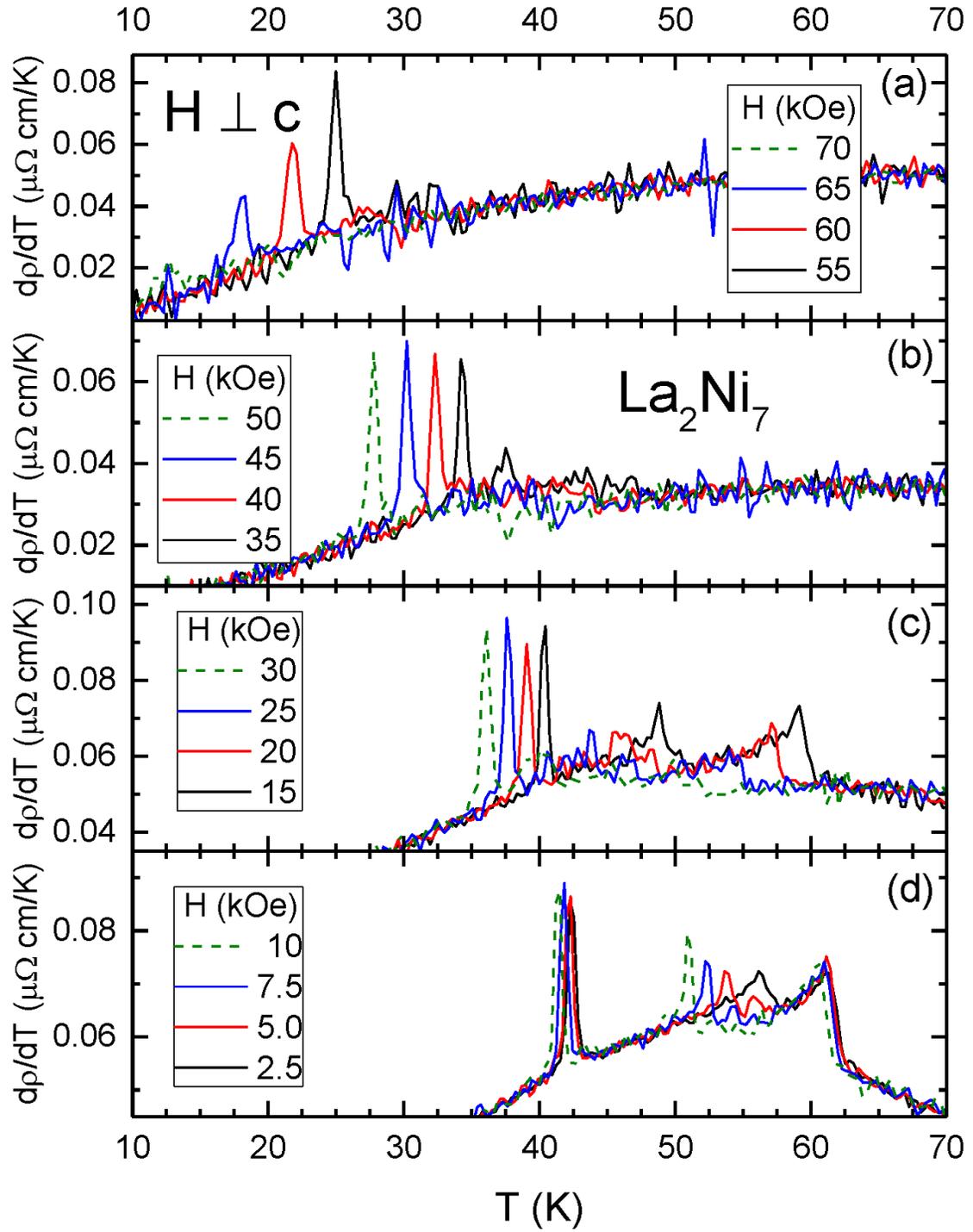

Fig. 17: dρ/dT plots for data shown in fig. 16.

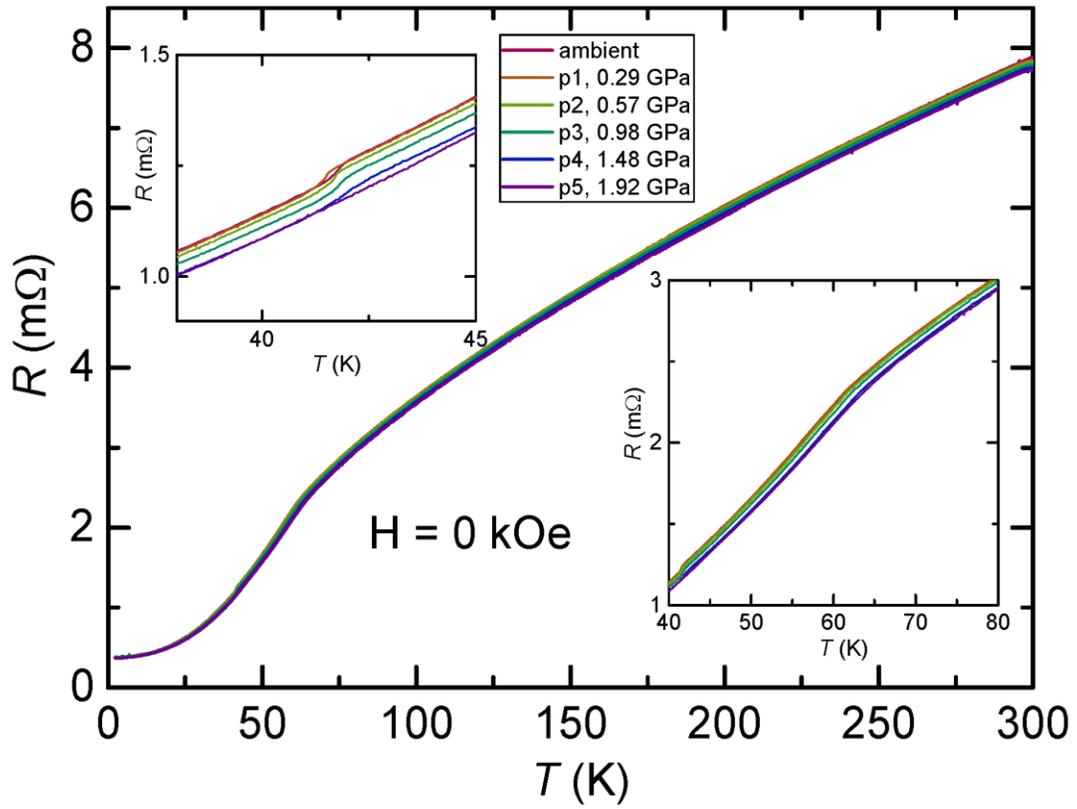

Fig. 18: Temperature dependent resistance, R(T), of $La_2Ni_7$ for applied pressures, p < 2.0 GPa; upper inset: expanded view centered on 42 K, lower inset: expanded view centered on 60 K.

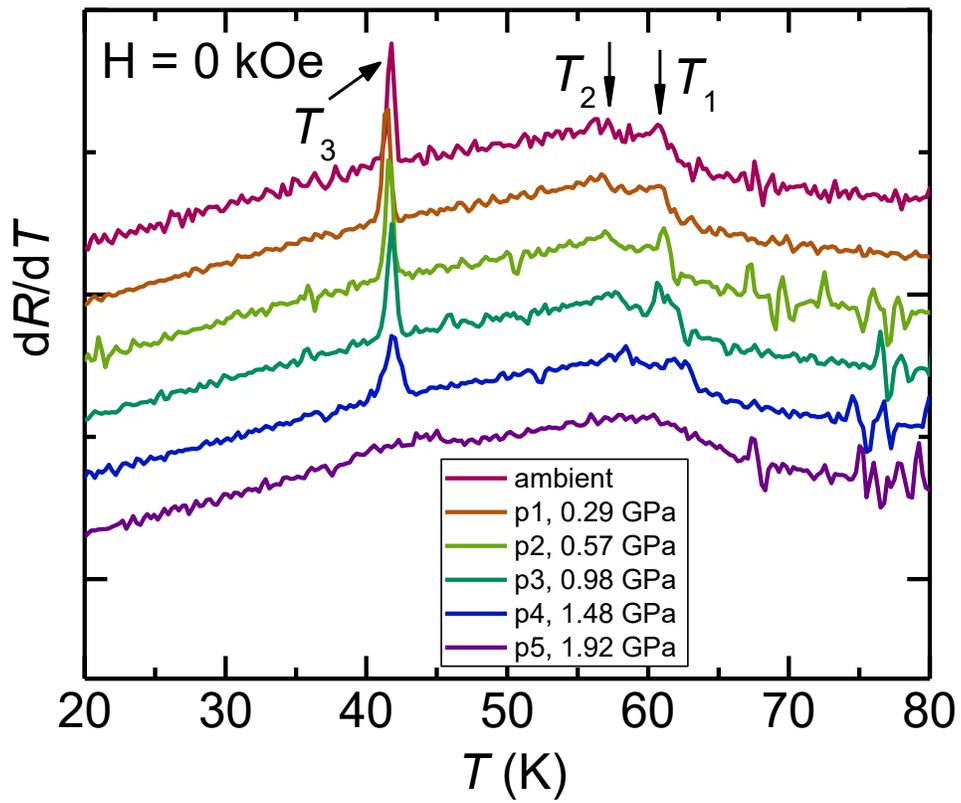

Fig. 19: dR(T)/dT plots for $La_2Ni_7$ under pressure for p < 2.0 GPa for 20 K < T < 80 K based on the data shown in fig. 18.

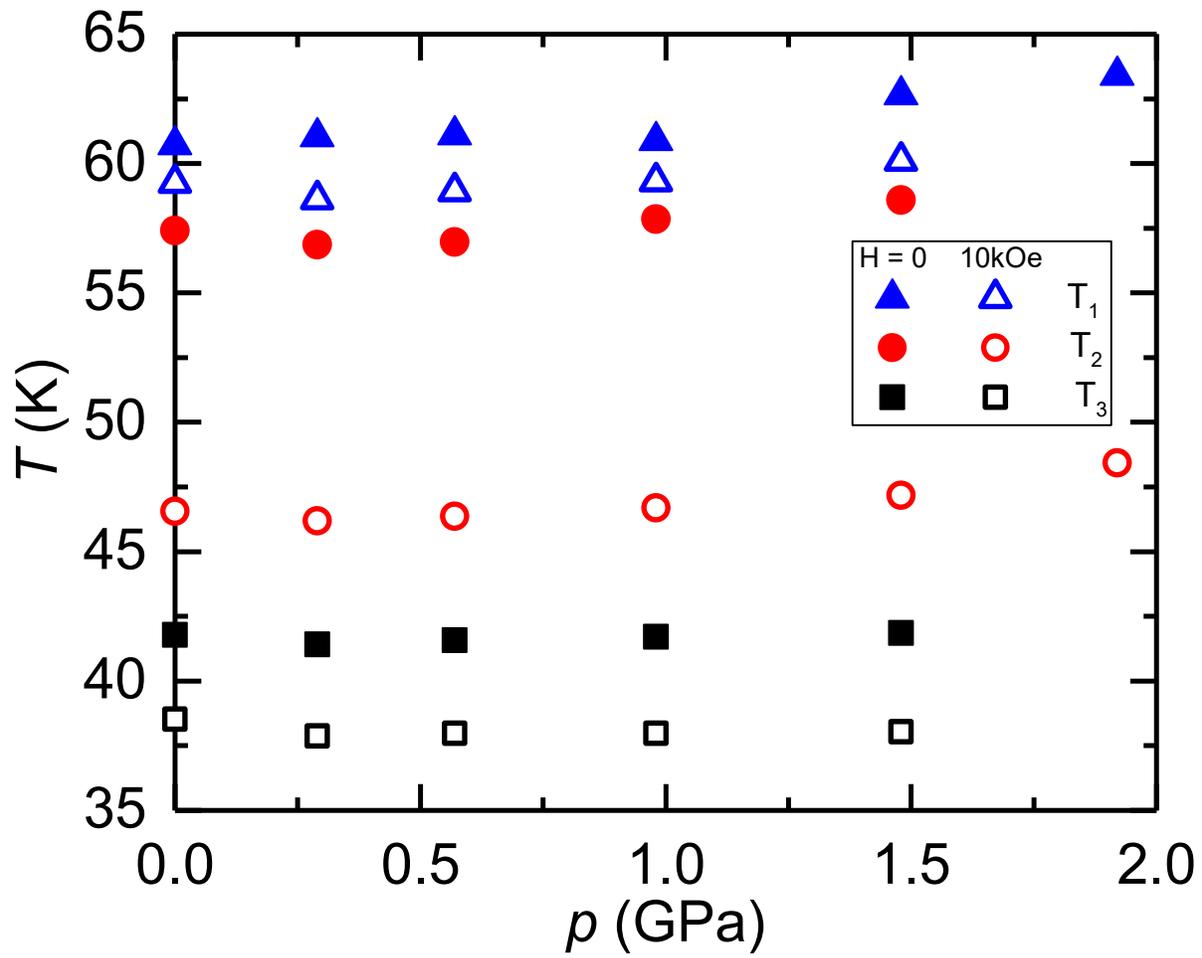

Fig. 20: T-p phase diagram for $La_2Ni_7$ for H = 0 (solid points) and $H_{\|c}$ = 10 kOe (open points).

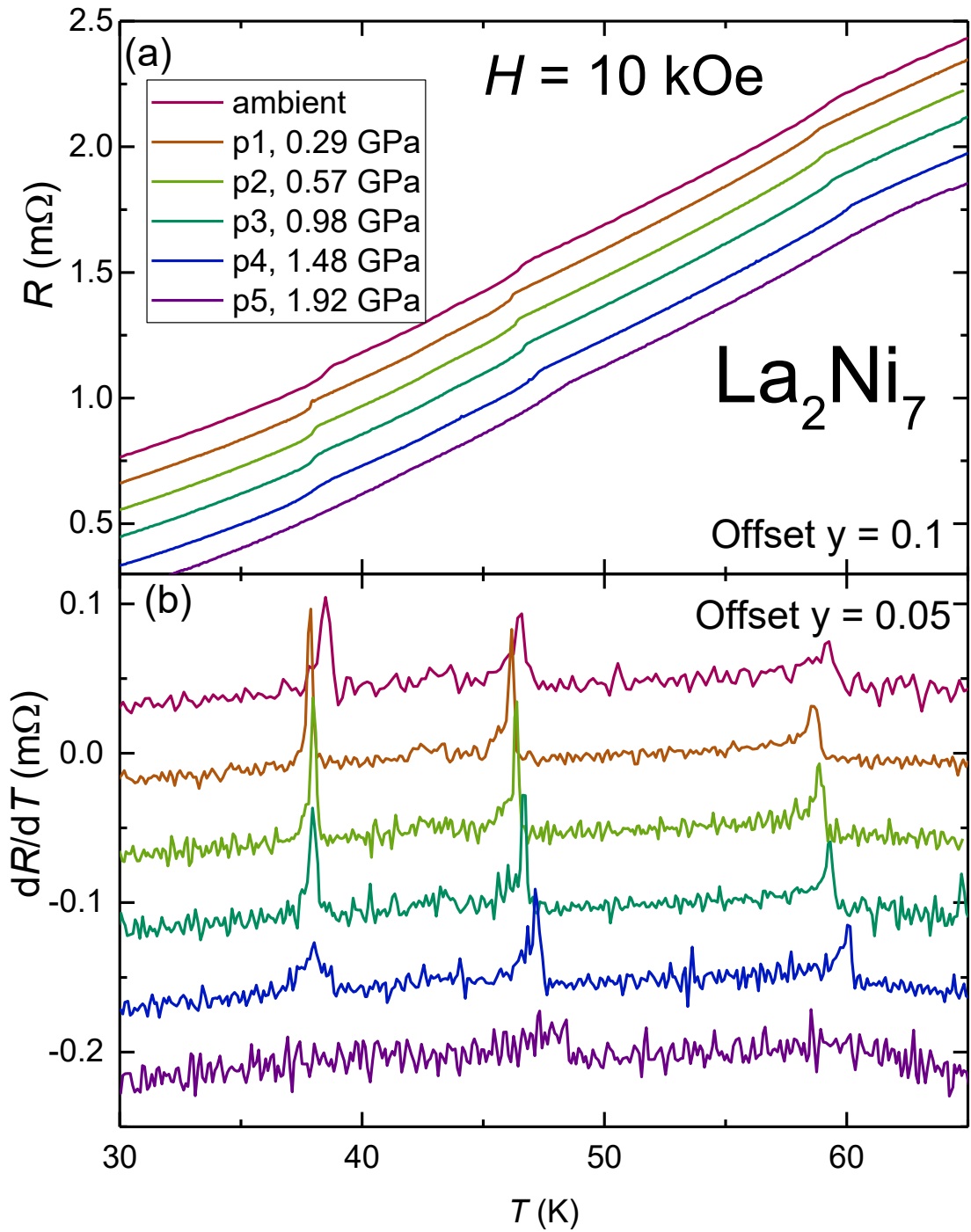

Fig. 21: (a) R(T) for $La_2Ni_7$ for 30 K < T < 65 K with a magnetic field of $H_{||c}$ = 10 kOe; data curves are offset from each for clarity. (b) dR/dT plots of the data shown in (a); data curves are offset from each for clarity.

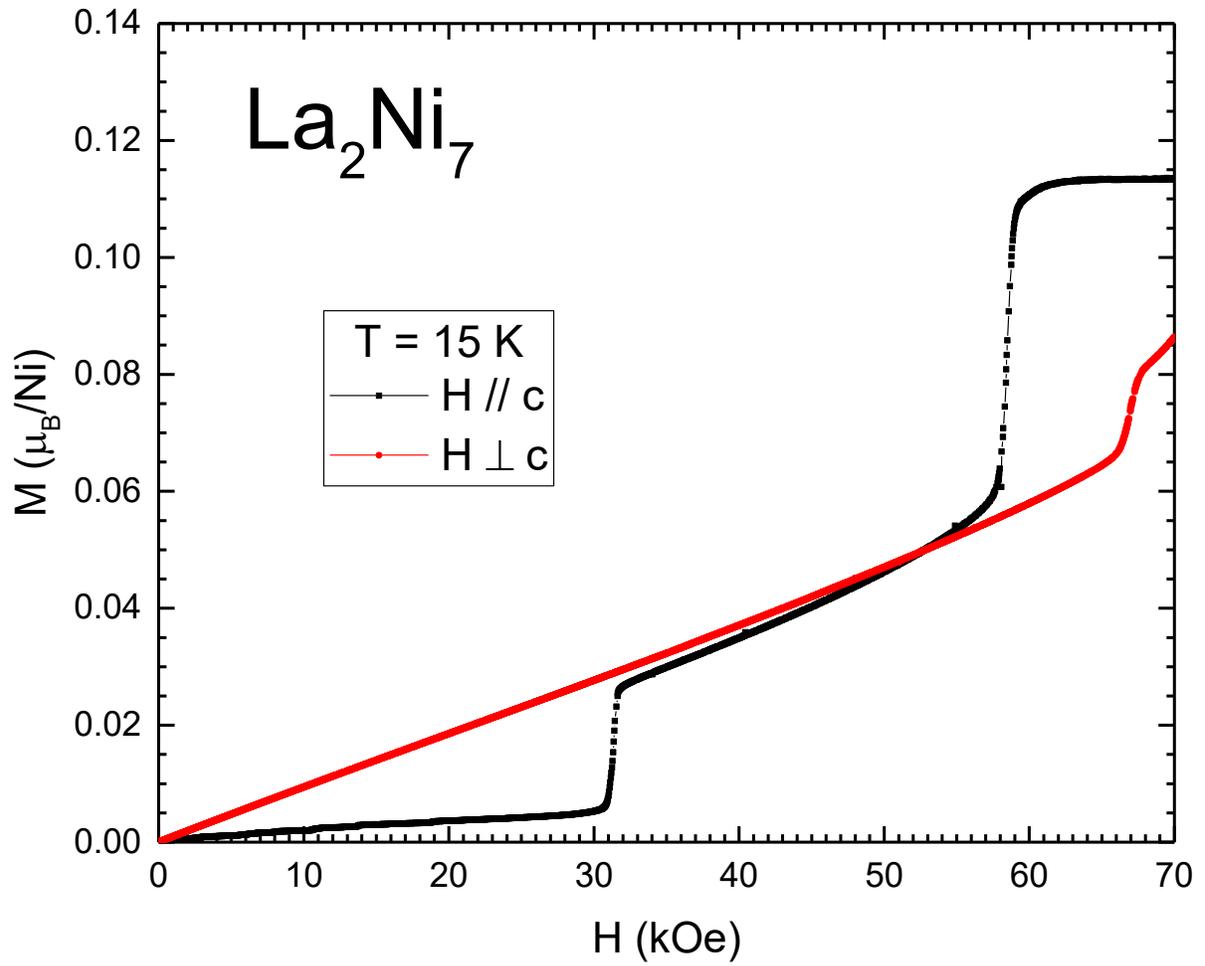

Fig. 22: Anisotropic, $H_{\parallel c}$ and $H_{\perp c}$, M(H) data for T = 15 K.

**References:**

[1] P. C. Canfield and S. L. Bud'ko, Rep. Prog. Phys. **79**, 084506 (2016).

[2] D. Belitz, T. R. Kirkpatrick, and T. Vojta, Phys. Rev. B **55**, 9452 (1997).

[3] A. V. Chubukov, C. Pepin, and J. Rech, Phys. Rev. Lett. **92**, 147003 (2004).

[4] G. J. Conduit, A. G. Green, and B. D. Simons, Phys. Rev. Lett. **103**, 207201 (2009).

[5] U. Karahasanovic, F. Krüger, and A. G. Green, Phys. Rev. B **85**, 165111 (2012).

[6] S. J. Thomson, F. Krüger, and A. G. Green, Phys. Rev. B **87**, 224203 (2013).

[7] C. J. Pedder, F. Krüger, and A. G. Green, Phys. Rev. B **88**, 165109 (2013).

[8] G. Abdul-Jabbar, D. A. Sokolov, C. D. O'Neill, C. Stock, D. Wermeille, F. Demmel, F. Krüger, A. G. Green, F. Levy-Bertrand, B. Grenier, and A. D. Huxley, Nat. Phys. **11**, 321 (2015).

[9] V. Taufour, U. S. Kaluarachchi, R. Khasanov, M. C. Nguyen, Z. Guguchia, P. K. Biswas, P. Bonfà, R. De Renzi, X. Lin, S. K. Kim, E. D. Mun, H. Kim, Y. Furukawa, C.-Z. Wang, K.-M. Ho, S. L. Bud'ko, and P. C. Canfield, Phys. Rev. Lett. **117**, 037207 (2016).

[10] U. S. Kaluarachchi, S. L. Bud'ko, P. C. Canfield, and V. Taufour, Nat Commun **8**, 546 (2017).

[11] E. Gati, J. M. Wilde, R. Khasanov, L. Xiang, S. Dissanayake, R. Gupta, M. Matsuda, F. Ye, B. Haberl, U. Kaluarachchi, R. J. McQueeney, A. Kreyssig, S. L. Bud'ko, and P. C. Canfield, Phys. Rev. B **103**, 075111 (2021).

[12] L. Xiang, E. Gati, S. L. Bud'ko, S. M. Saunders, and P. C. Canfield, Phys. Rev. B **103**, 054419 (2021).

[13] U. S. Kaluarachchi, L. Xiang, J. Ying, T. Kong, V. Struzhkin, A. Gavriliuk, S. L. Bud'ko, and P. C. Canfield, Phys. Rev. B **98**, 174405 (2018).

[14] U. S. Kaluarachchi, V. Taufour, S. L. Bud'ko, and P. C. Canfield, Phys. Rev. B **97**, 045139 (2018).

[15] A. V. Virkar and A. Raman, J. Less Comm. Met. **18**, 59 (1969).

[16] K. H. J. Buschow, J. Mag. Mag. Mater. **40**, 224 (1983).

[17] F. T. Parker, H. Oesterreicher, J. Less Comm. Met. **90**, 127 (1983).

[18] U. Gottwick, K. Gloos, S. Horn, F. Steglich, N. Grewe, J. Mag. Mag. Mater. **47&48**, 536 (1985).

[19] Y. Tazuke, R. Nakabayashi, S. Murayama, t. Sakakibara, T. Goto, Physica B **186-188**, 596 (1993).

[20] Y. Tazuke, M. Abe, S. Funahashi, Physica B **237-238**, 559 (1997).

[21] M. Fukase, Y. Tazuke, H. Mitamura, T. Goto, T. Sato, J. Phys. Soc. Japan **68**, 1460 (1999).

[22] M. Fukase, Y. Tazuke, H. Mitamura, T. Goto, T. Sato, Mater. Trans. JIM **41** 1046 (2000).

[23] Y. Tazuke, H. Suzuki, H. Tanikawa, Physica B **346-347**, 122 (2004).

[24] J-C. Crivello and V. Paul-Boncour, J. Phys.: Condens. Matter **32**, 145802 (2020).

[25] P. C. Canfield and I. R. Fisher, J. Cryst. Growth **225**, 155 (2001).

[26] P. C. Canfield, Rep. Prog. Phys. **83** 016501 (2020).

[27] A. C. Larson and R. B. Von Dreele, Los Alamos National Laboratory Report LAUR 86-748 (2000).

[28] B. H. Toby, J. Appl. Cryst. **34** 210 (2001).

[29] P. Villars, H. Okamoto, and K. Cenzual, ASM Alloy Phase Diagram Database Online, No. 1600322.

[30] H. Okamoto, J. Phase Equilib. **23**, 287 (2002).

[31] J. Dischinger and H. J. Schaller, J. Alloys Compd. **312**, 201 (2000).

[32] D. Zhany, J. Tang, K. A. Gschneidner, Jr., J. Less Comm. Metal **169**, 45 (1991).

[33] S. L. Bud'ko, A. N. Voronovskii, A. G. Gapotchenko, and E. S. Itskevich, Zh. Eksp. Teor. Fiz. **86**, 778 (1984).

[34] S. K. Kim, M. S. Torikachvili, E. Colombier, A. Thaler, S. L. Bud'ko, and P. C. Canfield, Phys. Rev. B **84**, 134525 (2011).

[35] M. S. Torikachvili, S. K. Kim, E. Colombier, S. L. Bud'ko, and P. C. Canfield, Rev. Sci. Instrum. **86**, 123904 (2015).

[36] B. Bireckoven and J. Wittig, J. Phys. E **21**, 841 (1988).

[37] L. Xiang, E. Gati, S. L. Bud'ko, R. A. Ribeiro, A. Ata, U. Tutsch, M. Lang, and P. C. Canfield, Rev. Sci. Instrum. **91**, 095103 (2020).

[38] M. E. Fisher and J. S. Langer, Phys. Rev. Lett. **20**, 665 (1968).

[39] M. E. Fisher, Philos. Mag. **7**, 1731 (1962).

[40] R. A. Ribeiro, S. L. Bud'ko, and P. C. Canfield, J. Magn. Magn. **267**, 216 (2003).

[41] T. A. Wiener and P. C. Canfield, Journal of Alloys and Compounds **303–304**, 505 (2000).

[42] T. Kong, C. Cunningham, V. Taufour, S. L. Bud'ko, M. L. C. Buffon, X. Lin, H. Emmons, and P. C. Canfield, J. Magn. Magn. **358-359**, 212 (2014).

[43] Rhodes, P.; Wohlfarth, E.P. The effective Curie-Weiss constant of ferromagnetic metals and alloys. Proc. R. Soc. Lond. A **273**, 247 (1963).

[44] J. P. Perdew, K. Burke, and M. Ernzerhof, Physical Review Letters **77**, 3865 (1996).

[45] Although PBE values of compressibility are likely smaller than experimental values, for relative comparison it can provide some insight. The values found were 115, 88 and 45 GPa for $La_2Ni_7$, $LaCrGe_3$ and $EuCd_2As_2$ respectively.

[46] E. Gati, S. L. Bud'ko, L. Wang, A. Valadkhani, R. Gupta, B. Kuthanazhi, L. Xiang, J. M. Wilde, A. Sapkota, Z. Guguchia, R. Khasanov, R. Valenti, and P. C. Canfield, arXiv:2108.01871.

[47] Bhattacharyya A, Jain D, Ganesan V, Giri S and Majumdar S   Investigation of weak itinerant ferromagnetism and critical behavior of $Y_2Ni_7$ Phys. Rev. B **84** 184414 (2011).

[48] L. Holmes, M. Eibschutz, and H. J. Guggenheim, Solid State Communications **7**, 973 (1969).

[49] T. Nagamiya, K. Yosida, and R. Kubo, Adv. Phys. **4**, 1 (1955).